\RequirePackage[2020-02-02]{latexrelease}
\documentclass[aps,prl,twocolumn,groupedaddress,floatfix]{revtex4-2}

\usepackage{mathtools}
\usepackage{ifsym}
\usepackage{wasysym}
\usepackage{graphicx}
\usepackage{siunitx}
\newcommand{\rg}{R_\text{g}}
\newcommand{\rh}{R_\text{H}}
\usepackage[dvipsnames]{xcolor}

\begin{document}

\title{Numerical Study of Neutral and Charged Microgel Suspensions: \\From Single-Particle to Collective Behavior}

\author{Giovanni Del Monte}
\affiliation{Soft Condensed Matter and Biophysics, Debye Institute for Nanomaterials Science, Utrecht University, Utrecht, Netherlands}
\affiliation{CNR-ISC and Department of Physics, Sapienza University of Rome, p.le A. Moro 2, 00185 Roma, Italy}
\author{Emanuela Zaccarelli}
\email{emanuela.zaccarelli@cnr.it}
\affiliation{CNR-ISC and Department of Physics, Sapienza University of Rome, p.le A. Moro 2, 00185 Roma, Italy}

\date{\today}

\begin{abstract}
We perform extensive Molecular Dynamics simulations of an ensemble of realistic microgel particles in swollen conditions in a wide range of packing fractions $\zeta$. We compare neutral and charged microgels, where we consider charges distribution adherent to experimental conditions. Through a detailed analysis of single-particle behavior, we are able to identify the different regimes occurring upon increasing concentration: from shrinking to deformation and interpenetration, always connecting our findings to available experimental observations. We then link these single-particle features to the collective behavior of the suspension, finding evidence of a structural reentrance, that has no counterpart in the dynamics. Hence, while the maximum of the radial distribution function displays a non-monotonic behavior with increasing $\zeta$, the dynamics, quantified by the microgels' mean-squared displacement, always slows down. This behavior, at odds with simple Hertzian model, can be described by a phenomenological multi-Hertzian, which takes into account the enhanced internal stiffness of the core. However, also this model fails when deformation enters into play, whereby more realistic many-body models are required.  Thanks to our analysis, we are able to unveil the key physical mechanisms, shrinking and deformation, giving rise to the structural reentrance that holds up to very large packing fractions. We further identify key similarities and differences between neutral and charged microgels, for which we detect at high enough $\zeta$  the fusion of charged shells, previously invoked to explain key experimental findings, and responsible for the structural reentrance. Overall, our study establishes a powerful framework
to uncover the physics of microgel suspensions, paving the way to tackle different regimes, e.g. high temperature, and internal architectures, such as for hollow and ultra-low-crosslinked microgels, where experimental evidence is still limited.
\end{abstract}

\maketitle

\section{Introduction}

Nowadays, a variety of colloidal particles can be synthesized, with different shape, internal composition and/or mutual interactions~\cite{glotzer2007}. This versatility provides the possibility to control each of these properties at the single-particle level in the ultimate aim to transfer them to the macroscopic scale through a bottom-up approach~\cite{zhang2003building}. However, this last step is not always easy to perform, because it is often difficult to simultaneously obtain detailed information at both microscopic and macroscopic length scales. Thus, there is a fundamental need for investigations able to bridge such scales in order to finally predict the behavior of materials from the features of individual building blocks. 

In this context, single-particle resolution experiments come at hand. This type of studies in colloidal suspensions started with confocal microscopy experiments which opened up the possibility of studying static and dynamic behavior of concentrated systems with single particle precision. For example, pioneering works revealed the role of dynamic heterogeneities in hard-sphere colloids close to the glass transition~\cite{weeks2000three} or the occurrence of arrested phase separation in depletion gels~\cite{lu2008gelation}. More recently, the spreading of super-resolution microscopy~\cite{voetssuper} has enhanced the spatial resolution reachable in experiments, 
providing a new important tool which can be used either to extend the observation timescale for glassy dynamics~\cite{hallett2018local} or to enable the visualization of particles inner structure~\cite{bergmann2018super,otto2020resolving} and functionalization~\cite{pujals2019super}.

Among the tunable features of colloidal particles, a very important one is softness~\cite{vlassopoulos2014tunable}, which can be defined as the ratio between elastic free energy and thermal energy. Using colloids of polymeric nature, several orders of magnitude in softness can be spanned, ranging from chains to star polymers, microgels and more complex macromolecules. It thus becomes possible to investigate concentrated states well above single particle overlap. Among soft particles, microgels are colloidal-scale polymer networks dispersed in a solvent, that recently emerged as favourite model systems both for fundamental studies~\cite{yunker2014physics,scheffold2020} and for a variety of applications~\cite{karg2019,farooqi2017temperature,kittel2022translating,agrawal2018functional}. Indeed,
microgels can be made responsive to external stimuli, among which the easiest one to control is temperature. For example using a thermoresponsive polymer, such as poly(N-isopropylacrylamide) (pNIPAM), a Volume Phase Transition is observed from swollen to collapsed microgels upon increasing temperature.   This is an echo of the coil-to-globule transition of the corresponding linear chains, arising as a consequence of the change in polymer-solvent affinity as temperature increases. A recent review has focused on the different experimental techniques nowadays available to quantify softness in microgels~\cite{scotti2022softness}, highlighting the growing interest in these systems. 

Upon increasing concentration, microgels undergo different regimes, where they shrink, deform or  interpenetrate, forming ultradense states that have no counterpart in hard-sphere based colloids. To tackle these states, several investigations have been conducted in dense conditions to assess the properties of the overall suspensions, in particular by rheology~\cite{scheffold2010brushlike,pellet2016glass}, imaging~\cite{bouhid2017deswelling} and Dynamic Light Scattering~\cite{philippe2018glass}. To gain information on the single particle behavior, confocal microscopy~\cite{paloli2013fluid} as well as a variety of neutron scattering techniques~\cite{cors2018determination} has been employed. For example, in the case of responsive microgel particles, that are the focus of this work, a clever contrast-matching (Zero-Average Contrast, ZAC)~\cite{mohanty2017interpenetration} or selective deuteration of particles~\cite{keerl2009temperature,scotti2019deswelling,zhou2023measuring} were adopted, in order to indirectly determine the changes to their internal structure upon crowding.  These measurements are nowadays complemented by a growing number of super-resolution microscopy investigations~\cite{conley2017jamming,bergmann2018super,conley2019}.

Despite the abundance of experimental measurements, simulation works devoted to described microgel collective behavior have been so far quite scarce. Most of them were based on simple coarse-grained effective models~\cite{rovigatti2019numerical}, starting with the widely used Hertzian model. This two-body potential, calculated for elastic spheres at low deformation, is known to display a so-called reentrant dynamical behavior with increasing packing fraction~\cite{berthier2010increasing}: the (self or collective) particle relaxation initially slows down and then becomes faster again upon crowding.
This counter-intuitive behavior has not been reported in microgels suspensions~\cite{philippe2018glass,conley2019}, although the situation for microgels adsorbed at liquid-liquid interfaces may be interestingly different~\cite{camerin2020microgels,schmidt2023}.  This lack of non-monotonic dynamics clearly suggests that the Hertzian potential is too soft to be representative of microgels under dense conditions, as also confirmed by explicit interaction potential calculations~\cite{rovigatti2019connecting}. Thus, additional contributions, describing the complex  internal stucture of real microgels, should be employed to improve the description. Among the proposed possibilities to go beyond the Hertzian model, we mention the use of the phenomenological multi-Hertzian model~\cite{bergman2018new}, the incorporation of isotropic volume changes in the Hamiltonian~\cite{urich2016swelling,baul2021structure} or the development of more complex toy-models including internal elasticity~\cite{gnan2021dynamical,boattini2020modeling}. 
\begin{figure*}[ht]
	\centering
	\includegraphics[width=0.75\linewidth]{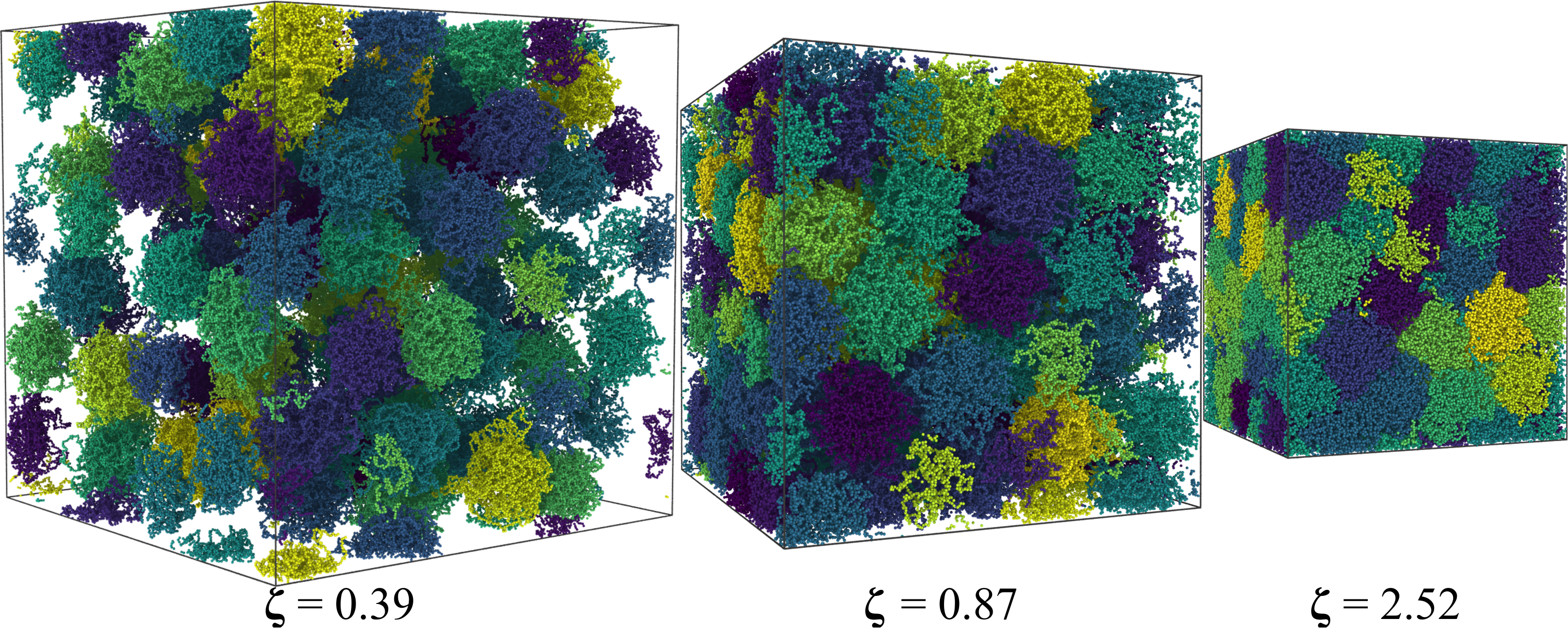}
	\caption{Representative snapshots of neutral microgel suspensions at packing fractions $\zeta=0.39, 0.87$ and $2.52$, from left to right, respectively. Box dimensions are reproduced with the right proportions, so that deswelling of the microgels can be visualized.}
		\label{fig:snap}
\end{figure*}

A different approach is instead to directly simulate realistic models of microgels at different concentrations, which is of course computationally more challenging.  Only a recent study of this kind exists~\cite{nikolov2020behavior}, where extensive simulations of microgels were performed with the aim to establish a relationship between the bulk modulus of the individual particles and that of the whole suspension under different density and solvent conditions. However, the performed study lacked a full control on the internal structure of the simulated microgels, consisting of homogeneous chain assemblies.
Conversely, we recently developed a realistic microgel model~\cite{gnan2017silico}, which at the individual level is able to fully capture the typical core-corona structure of PNIPAM microgels synthesized by precipitation polymerization, that are the ones mostly used in the experiments. We  showed that the model is robust to crosslinker concentration~\cite{hazra2023structure} and  system size~\cite{ninarello2019modeling} variation, hence also microgels with few thousands monomers each are able to reproduce the main features that we aim to incorporate in the simulations. 
Another important point that was overlooked in Ref.~\cite{nikolov2020behavior} is that pNIPAM microgels always carry an intrinsic charge, due to the presence of ionic initiators in the synthesis. This charge is mostly screened by the counterions under dilute and swollen conditions, but several studies have highlighted the role of counterions-induced osmotic pressure on the behavior of concentrated suspensions~\cite{scotti2016role,del2021two,zhou2023measuring,zhou2023poly}. 

The present work fills the gap by performing extensive Molecular Dynamics (MD) simulations of an ensemble of realistic microgels under swollen conditions in an extended range of packing fractions. For the first time, we report numerical results of suspensions of charged microgels and carefully compare results for neutral and charged systems, highlighting the differences between them. Our study provides an extensive characterization of both individual structural properties and collective behavior as a function of increasing microgel concentration. We are thus able to address the onset and relative interplay of deswelling, faceting and interpenetration under different concentration regimes and find quantitative agreement with available super-resolution experiments. We also calculate individual microgel elasticity and connect it to the behavior of the whole suspension. We then examine in detail how the collective structure and dynamics are affected by crowding, again validating our model against experimental measurements available in the literature, clearly detecting the onset of structural reentrance, but not of a dynamical one. Both phenomena occur independently of the presence of charges, but with important differences in the microgel behavior when the latter are taken into account. Such decoupling of the reentrance between statics and dynamics was previously unreported and we explain it here with the help of a multi-Hertzian model, which allows us to grasp the physical motivations of such behavior. We are thus able to connect the physical mechanisms responsible for structural reentrance, thereby providing a novel, comprehensive picture of the single-particle and collective behavior of concentrated microgel suspensions. Our work offers a clear framework to robustly interpret experimental data and to capture a minimal-level description of the complex phenomenology of these fascinating systems.

\section{Results}

We start by reporting snapshots of the system in Fig.~\ref{fig:snap} for neutral microgels at three studied packing fractions $\zeta$, as defined in Eq.~\ref{eq:packfrac}: below random close packing, just above it and at a very high one. These are representative of the typical states encountered in the experimental phase diagram of the system: fluids, crystals\footnote{Importantly, we stress that crystallization is often difficult to detect in simulations, due to the finite system size or to the long nucleation times, also disfavoured by the polydispersity of the employed microgels, but we will return to this point at the end of the manuscript.} and glasses~\cite{paloli2013fluid}, in analogy to hard sphere systems but occurring at much larger volume fractions due to the softness of the interactions.
In our simulations, we find that, as soon as the packing fraction increases above random close packing, the microgels get in contact and start to accommodate themselves in order to more efficiently fill the space. This is accomplished through several different mechanisms, involving volume or shape change and, alternatively,  interpenetration with other microgels. For charged microgels, similar mechanisms occur, with the main difference that non-negligible electrostatic interactions are found to play a role at all $\zeta$, including low ones.

\subsection{Individual microgel size and structure}
We initially focus on reporting the behavior of individual microgels, particularly by looking at the concentration dependence of their size  and at the evolution of the internal single-particle structure quantified by the form factors (see Methods).
This is a useful starting point that allows us to directly connect with small-angle neutron scattering experiments performed on hydrogenated microgels dispersed in a sea of deuterated ones~\cite{scotti2019deswelling,zhou2023measuring}. 
\begin{figure}[ht]
	\centering
	\includegraphics[width=0.85\linewidth]{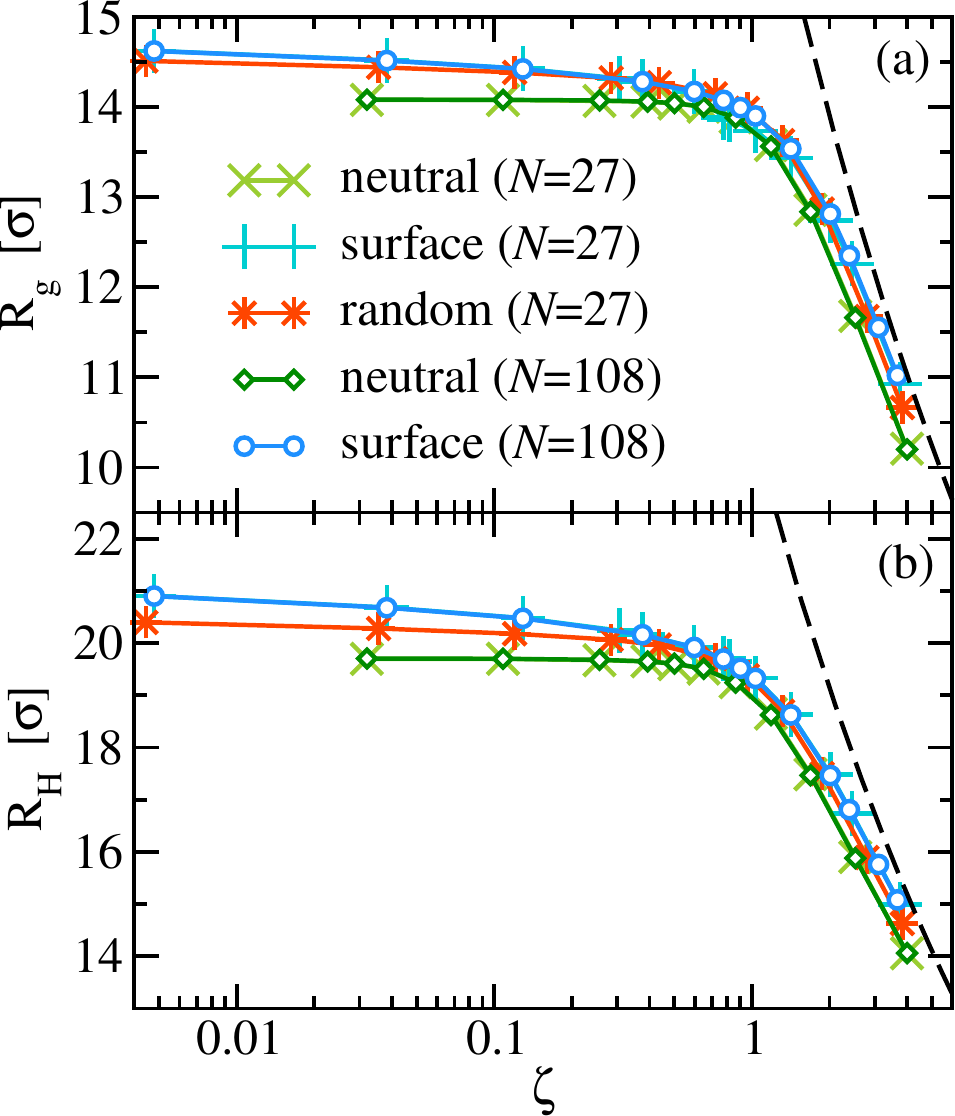}
	\caption{Gyration $\rg$ (a) and hydrodynamic $\rh$ (b) radii
	as a function of  packing fraction $\zeta$ for neutral and charged microgels, having both a random and surface charge distribution.	Data for two system sizes with $N=27$ and $N=108$ microgels are compared, showing  no detectable size effects. Dashed lines indicate isotropic shrinking behavior: $R_g, R_H\sim \zeta^{-1/3}$.}
	\label{fig1}
\end{figure}
To have an estimate of the microgels size as a function of packing fraction $\zeta$, we calculate the gyration radius $R_g$ and estimate the hydrodynamic radius $R_H$, both reported in Fig.~\ref{fig1} for neutral and charged microgels.
In the figure, we also compare data obtained with a different number of microgels ($N=27$ and $N=108$) and for two different charge distributions: a random one, where charges are evenly located throughout the whole microgel, and a surface one, where they are positioned on the microgel external surface~\cite{del2021two}, intended as more representative of the location of the ionic initiators after pNIPAM-based microgel synthesis.

For neutral microgels both $R_g$ and $\rh$ are found to be constant up to particle contact, corresponding to random close packing ($\zeta\sim 0.74$).
This clearly indicates no particle deswelling at low packing fractions, contrarily to what happens in charged microgels, irrespective of charges distribution.
In this case, both observables slightly decrease already at small concentrations. Hence, although this is a minor effect of the order of 5\% of the overall size of the particles, the shrinking is clearly detectable in the presence of changes, 
due to a variation of the screening conditions induced by counterions with increasing microgels concentration~\cite{scotti2016role}.
Conversely, experimental works have reported contradictory results on this point: in some cases~\cite{mohanty2017interpenetration,scotti2019deswelling} a flat behavior of the particle size is reported, similarly to the neutral simulations. In particular, Ref.~\cite{mohanty2017interpenetration} investigates microgels in the presence of added salt, which is likely to be more similar to the neutral case examined here.
Other studies instead clearly report deswelling even at low $\zeta$~\cite{zhou2023measuring} as for ionic ones. In addition, a recent work on co-polymerized microgels containing pNIPAM~\cite{ruiz2023concentration} has detected quite a significant shrinking happening well below particle contact.  
Focusing on larger values of $\zeta$, we observe a much more significant shrinking in all cases. Above contact, this is  compatible with isotropic shrinking, i.e. as $\sim \zeta^{-1/3}$.  The dependence on $\zeta$ for both $R_g$ and $\rh$ indicates that, despite  corresponding to quite different measures of the sizes of the microgels, they undergo a similar behavior as packing fraction increases. This is confirmed by the data obtained for different system sizes ($N=27$ and $N=108$), which clearly show no dependence at all for the individual microgel size at all studied $\zeta$. From now on, unless otherwise stated, all results refer to simulations with $N=108$ microgels. In addition, we notice that the random case is always intermediate between neutral and surface ones, hence from now on our more extended analysis will only focus on the latter two cases.
\begin{figure}[ht]
	\centering
	\includegraphics[width=0.85\linewidth]{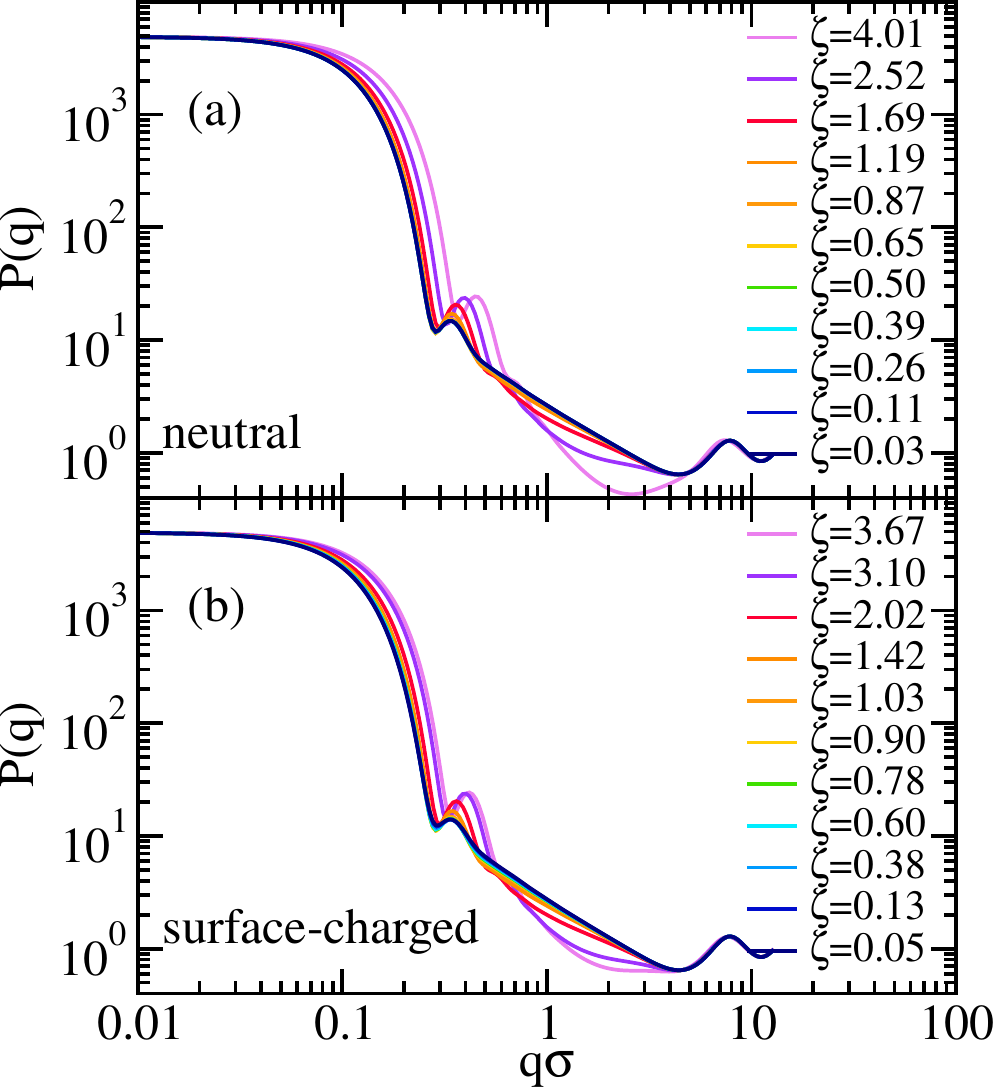}
	\caption{Average form factors $P(q)$ for $N=108$ (a) neutral and (b) surface-charged microgels at different packing fractions $\zeta$.}
	\label{fig:form-factors}
\end{figure}

In order to make a more direct connection with experiments, we next calculate the average form factors $P(q)$ of the individual microgels as a function of $\zeta$, reported in Fig.~\ref{fig:form-factors} for neutral and surface-charged microgels. In both cases, $P(q)$ remains rather unchanged with increasing $\zeta$ up to effective packing fractions close to 1. For larger $\zeta$, the first peak starts to move to larger $q$ values and to increase in height, with noticeable changes in the slope of data after the first peak. In this respect, the form factors start to resemble those of more collapsed microgels, possibly also deformed, as we will see later on.

\begin{figure}[ht]
	\centering
	\includegraphics[width=0.85\linewidth]{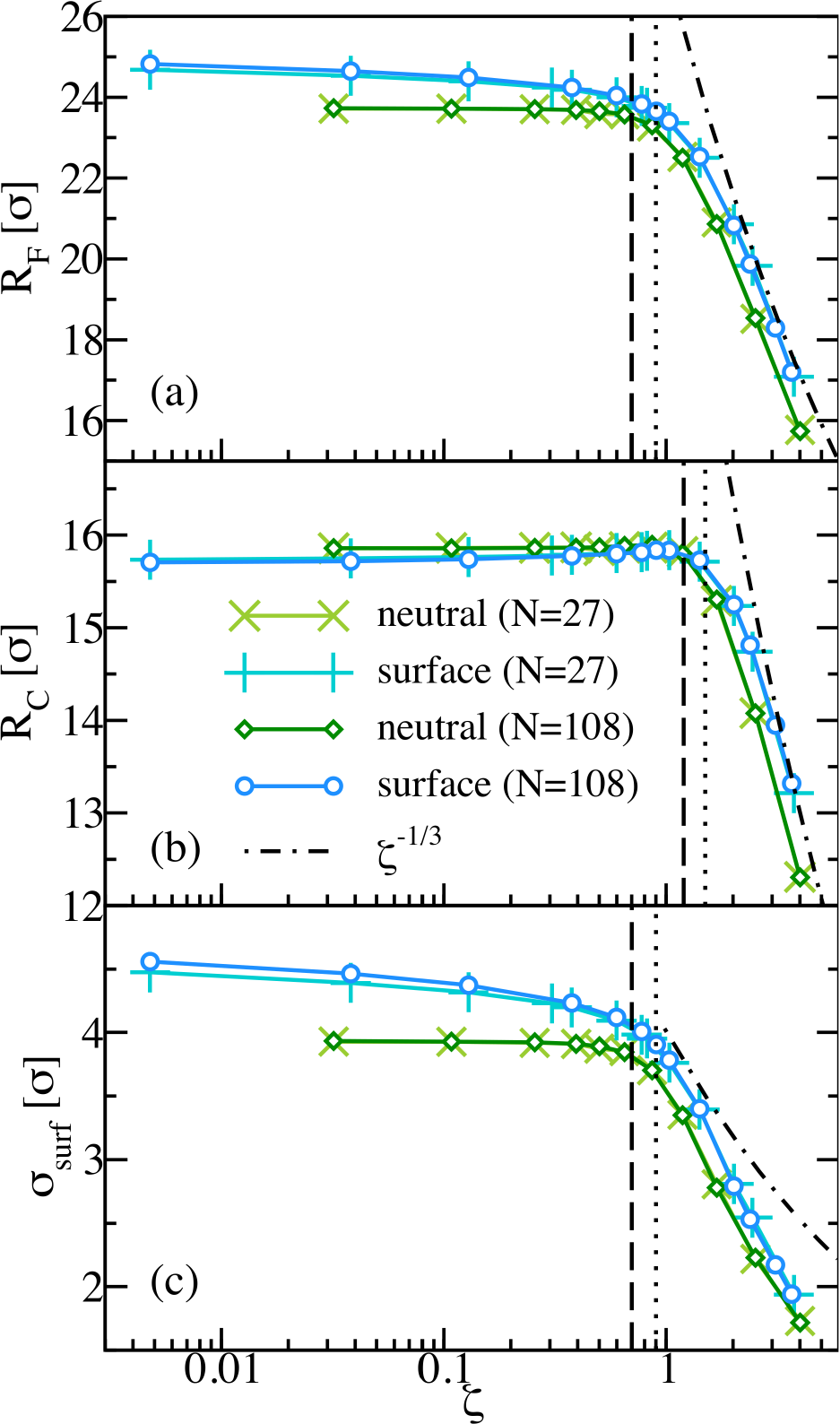}
	\caption{Parameters extracted from the fits of the form factors with Eq.~\ref{eq:mod_fuzzy} as a function of $\zeta$: (a) total radius $R_F$,  (b) core radius $R_C$ and (c) half of fuzzy shell thickness $\sigma_\text{surf}$. Data corresponding to two system sizes are reported:  $N=27$ and $N=108$. The  vertical lines in panels (a) and (c) represent the approximate start of isotropic shrinking for neutral microgels ($\zeta\sim 0.7$, dashed lines) and for charged ones ($\zeta\sim 0.9$, dotted lines), while the ones in panel (b) denote the start of the core shrinking, roughly taking place for $\zeta \gtrsim 1.2$ for neutral (dashed) and $\zeta \gtrsim 1.5$ for surface (dotted) microgels. Dashed-dotted lines at high $\zeta$ represent the expected isotropic shrinking behavior ($\sim \zeta^{-1/3}$). The extension of the shell in panel (c) is found to decrease much faster at very large $\zeta$.}	\label{fig:fuzzy-r-vs-z}
\end{figure}

To compare our results with available SAXS or SANS experiments, we fit the form factors with the extended fuzzy sphere model (Eq.~\ref{eq:mod_fuzzy}), as described in Methods. In this way, we extract an estimate of the total radius $R_F$, as well as of the core radius $R_C$  and of the half-width of the fuzzy shell $\sigma_\text{surf}$. All of these observables are reported as a function of $\zeta$ in Fig.~\ref{fig:fuzzy-r-vs-z}. The fitting procedure has been repeated for simulations of systems with both $N=27$ and $N=108$ microgels and again we find that the results are very similar for both sizes, so that we can consider our simulation size appropriate for this kind of study. 
Panel (a) shows results for $R_F$, which is completely constant below particle contact for neutral microgels, in agreement with the findings of Fig.~\ref{fig1}, while a small decrease is observed for charged ones.
In both systems, a more pronounced shrinking starts to occur above particle contact, which happens at slightly higher packing fractions for charged microgels. The observed deswelling is again compatible with isotropic shrinking, similarly to what observed for $R_g$ and $R_H$.
Panel (b) displays the size of the core separately, where the core radius $R_c$ is shown to remain constant up to much larger values of $\zeta$, exceeding particle contact, for both neutral and charged microgels. Then, isotropic shrinking of the cores is found to occur only for $\zeta\gtrsim 2$.
Instead, the external shell, whose extension is quantified by $\sigma_\text{surf}$, starts decreasing earlier, just above contact. Evidently, the initial decrease of the microgel size around particle contact is only determined by the shrinking of the shell.
Only later, the cores come into play. This is also reflected in the behavior of $\sigma_\text{surf}$, which is not found to be compatible with an exponent $-1/3$ at all $\zeta$, but only in its initial shrinking trend.
Later on, when isotropic core shrinking happens, the shell is found to decrease even more, with a behavior roughly compatible with an exponent of $\approx -1/2$. This may be due to the softer character of the corona, that at extremely large packing fractions becomes completely deformed and collapses onto the core. 

\subsection{Individual microgel deformation}
\begin{figure}[ht]
	\centering
	\includegraphics[width=0.95\linewidth]{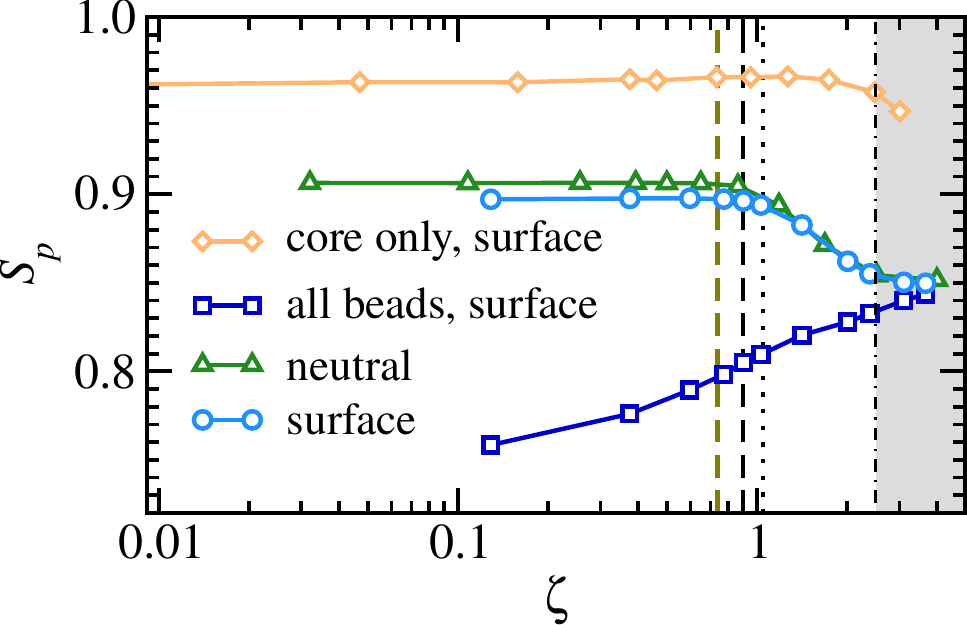}
	\caption{ Deformation parameter $S_p$ as a function of nominal packing fraction $\zeta$ for neutral and surface-charged microgels, considering all beads (surface only), only beads within $l_h=R_c+\sigma_\text{surf}$ (neutral and surface) and only core beads (surface). The vertical lines represent different thresholds discussed in the text: contact (dashed, $\zeta\sim 0.7$),  start of deformation for neutral (dashed, $\zeta\sim 0.9$) and for surface-charged microgels (dotted, $\zeta\sim 1.1$); end of deformation  (dot-dashed and grey region, $\zeta\sim 2.2$). 
 }
	\label{fig:deformation}
\end{figure}

Having evaluated the individual particle size as a function of $\zeta$, quantifying the isotropic deswelling that takes place with increasing concentration, we now focus on the assessment of particle deformation.
To this aim, we evaluate the deformation parameter $S_p$, defined in Eq.~\ref{eq:deformation}, that is an adimensional function of microgel volume and surface that is equal to 1 for a sphere and lower for other geometries.
This quantity is the 3D analogue of the ratio used in super-resolution experiments of Ref.~\cite{conley2017jamming}.
First of all, we calculated $S_p$ for each microgel considering all beads, as reported in Fig.~\ref{fig:deformation}, and with this procedure find the counterintuitive result that the microgels appear to be more anisotropic at low than at high $\zeta$. This is due to the presence of the dangling chains around the microgel, which provide an important contribution to the volume when the microgels are isolated, thus making them fluctuate a lot in shape and being instantaneously (but not on average) rather anisotropic.
Instead, when they get in contact with each other, the external chains  deform very easily and hence the microgels retain a more spherical shape. The enhanced sphericity is signalled by the increase of $S_p$, which, however, is always found to be much smaller than 1.
This picture contradicts the standard expectations, as well as experimental observations~\cite{bouhid2017deswelling,conley2019}, on particle deformation of concentrated microgels at high $\zeta$, due to the fact that such external chains cannot be easily detected in experiments, given their very low density.
The role of dangling chains is a matter of debate~\cite{boon2017swelling,nojd2018deswelling}, not being easily quantifiable in experiments, while they can be visualized in simulation studies like the present one.
Turning back to the deformation assessment, the important insight that we obtained from these results is that the dangling chains contribution should not be considered to compare with microscopy experiments and, for this reason, we neglect the outer chains in the calculation of the deformation parameter, considering only a subset of the beads that are more representative of the size of each microgel. To find such an estimate, we rely on the knowledge obtained by the form factors and use the length $l_h=R_c+\sigma_\text{surf}$, corresponding to the core radius plus half of the fuzzy shell width, as a threshold length above which we discard the most external and very low density beads, to calculate the microgel volume.
We stress that we use the value of $l_h$ estimated in dilute conditions, keeping it constant at all $\zeta$, in order to be able to meaningfully detect the deformation of the coronas, that are gradually collapsing onto the core with increasing packing fraction.

Within $l_h$, we then calculate $S_p$ and find that its behavior qualitatively reproduces the experimental one, whose value is initially close to the spherical limit and stays constant up to and above particle contact ($\zeta \sim 0.7$ for neutral and $\zeta\sim 0.9$ for surface-charged microgels). Indeed, only after, we see that the anisotropy parameter starts to decrease, in an almost linear fashion in the range $1.1 \lesssim \zeta \lesssim 2.0$. This is then the strongest deformation regime, which ends roughly at $\zeta \sim 2.5$, above which $S_p$ reaches a final plateau. The behavior of $S_p$ is very similar for both neutral and charged microgels, apart from a slight shift in $\zeta$ of the different regimes. These results are once more in remarkable agreement with super-resolution experiments by Conley and coworkers~\cite{conley2017jamming}, who
also found three regimes for their 2D deformation data: a low-$\zeta$ plateau up to $\zeta \approx 1$, a linear decrease and then a high-$\zeta$ plateau for   $\zeta \gtrsim 2$. Similar behavior was also observed by Bouhid \textit{et al.}~\cite{bouhid2017deswelling} upon imaging concentrated microgels. Instead, the simulation results reported by Nikolov \textit{et al.}~\cite{nikolov2020behavior} were not able to identify the third high-$\zeta$ regime reported in the present work as well as in experiments. This was perhaps due to the fact that these authors used  a different variable to quantify anisotropy, namely the ratio between surface and volume, which cannot reach a plateau at high packing fractions. Indeed, this quantity was then found to display a minimum for $\zeta \sim 1$ that should correspond to the onset of deformation, but did not saturate at high  $\zeta$~\cite{nikolov2020behavior}. Therefore, we conclude that $S_p$ (or its 2D analogue used in Ref.~\cite{conley2017jamming}) is a more appropriate quantity to monitor individual particle deformation and capture the different experimental behaviors.

Having set a threshold for determining particle deformation, we then perform its calculation considering only the core beads, to get an assessment of the core deformation. This is also  reported in Fig.~\ref{fig:deformation} for the surface microgels only, due to the similarity between the two sets of data. We find that core deformation is practically absent everywhere, except for a very minor increase, signalled by a decrease of the relative $S_p$, for $\zeta \gtrsim 2.0$. Hence, we can conclude that this phenomenology is almost absent for microgels at the studied crosslinker concentration.

\begin{figure*}[t!]
	\centering
	\includegraphics[width=0.7\linewidth]{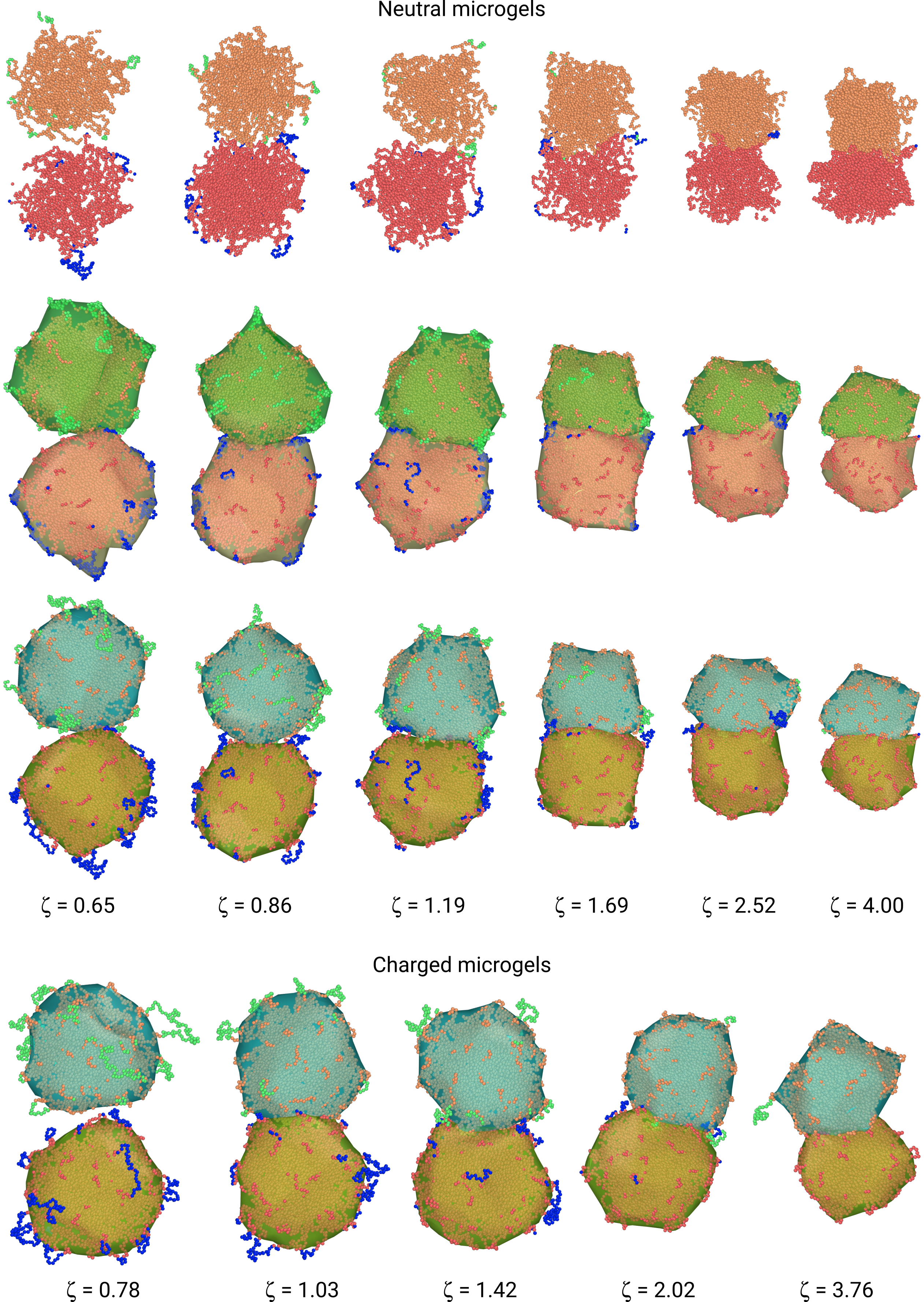}
	\caption{Snapshots representative of microgel neighbouring pairs at increasing $\zeta$ from left to right (see legends).  Top row: slice (to improve visualization) of neutral microgels with different colors distinguishing the inner beads considered in the deformation analysis (red and orange) and outer beads excluded from it (blue and green, whose distance from the microgel center of mass is greater than $l_h=R_c+\sigma_{\text{surf}}$); second row from top: surface mesh enclosing all the beads of the same microgels of the first row to highlight the anisotropic contributions of the outer beads especially at low packing fractions; third row from top:  surface mesh of the inner beads only, showing the volume variation of the microgels with increasing $\zeta$ that is quantified by the anisotropic parameters reported in Fig.~\protect\ref{fig:deformation}; fourth row: same as the third one but for surface-charged microgels.}
	\label{fig:shrinking}
\end{figure*}

\subsection{Microgel overlaps, faceting and effective packing fraction}
It is now instructive to look directly at the microgels to visually assess the progressive changes as $\zeta$ increases. To this aim, we report in Fig.~\ref{fig:shrinking}  representative pairs of neighbour microgels  at several different packing fractions. We start by discussing the neutral case for $\zeta$ just below contact (leftmost panels), where the microgels are not yet touching. In the top row, we report the full monomer representation, highlighting in a different color the dangling chains that we do not include in the deformation analysis discussed above, i.e. those that are found at a distance from the microgel center of mass greater than $l_h$. As it can be seen in Fig.~\ref{fig:shrinking}, these are very peripheral chains that protrude outside the microgel, therefore yielding deviations from the spherical shape at low $\zeta$. To better visualize this, in the second row we show the surface mesh of the microgels, calculating it over all beads including such outer chains, which highlights the asphericity of the surface close to these protrusions, thus explaining the low anisotropy parameter $S_p$ observed when considering all the beads in Fig.~\ref{fig:deformation}. Instead, when we exclude them from the calculation of the surface mesh, it is evident that the rest of the microgel is much more compact and overall spherical at low $\zeta$, as illustrated in the third row of Fig.~\ref{fig:shrinking}. Hence, we can now look at what happens with increasing $\zeta$: clearly, above contact the microgels start to shrink, maintaining their spherical-like shape and only later, for $\zeta \gtrsim 1.2$, they look evidently deformed and also start to facet. At this point, looking closely on the snapshots in the first row, it appears that along the sides where they are in close contact with their neighbours, they also start to interpenetrate, but this mechanism is never completely dominant. This may also be due to relatively large crosslinking ratio and small size of the investigated microgels, so that it will be interesting to compare this behavior with that of microgels having $c \sim 1\%$ or below in the future. 
\\
Turning to examine the surface-charged snapshots in the last row of Fig.~\ref{fig:shrinking}, it is clear that they remain overall larger in size at all studied packing fractions. The outer chains, as expected, are much more extended in this case, which further disfavour interpenetration. Indeed, such chains are more likely found to be within the interstitial space, as also visible in the snapshots. However, the degree of deformation of the microgels qualitatively appears to be similar to neutral case, in agreement with the behavior of $S_p$ in Fig.~\ref{fig:deformation}. 
\begin{figure*}[ht]
	\centering
	\includegraphics[width=0.9\linewidth]{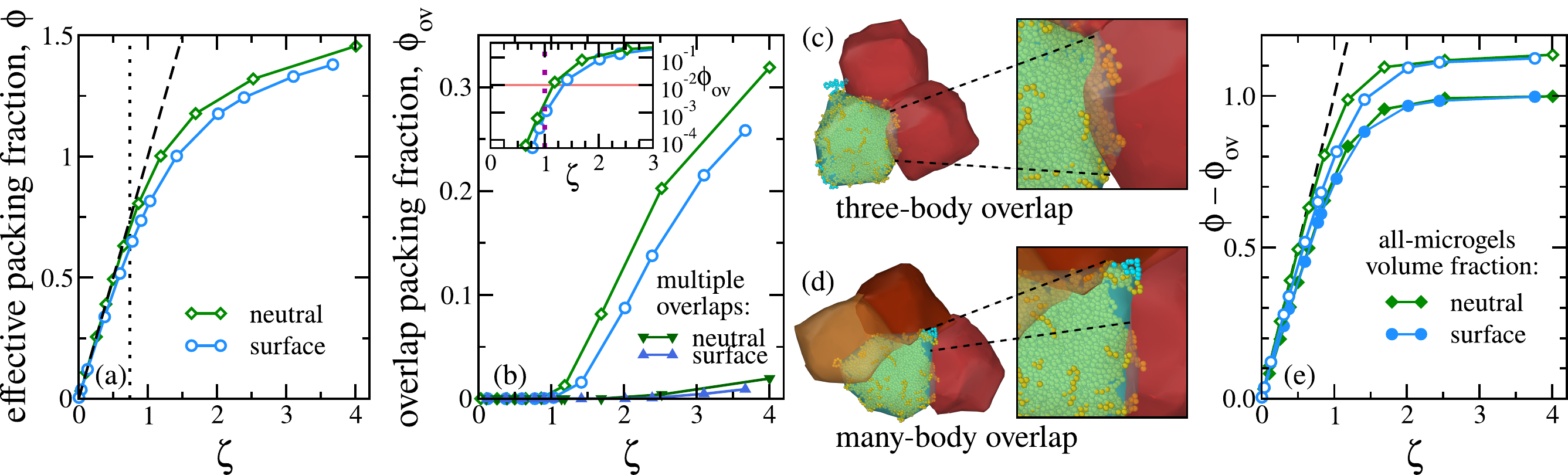}
	\caption{
 (a) effective packing fraction $\phi$, calculated from the hydrodynamic radius of the microgels at each state point, as a function of nominal packing fraction, $\zeta$, calculated using the hydrodynamic radius in dilute conditions; the black dashed line indicates $\phi=\zeta$ points, while the vertical dotted line denotes particle contact at $\zeta \sim 0.7$ where deviations from this behavior start to take place for neutral microgels.
 (b) Overlap packing fraction $\phi_{ov}$ as a function of $\zeta$, where the additional triangles represent the volume fraction for which three or more microgels overlap; the inset shows an enlargement of the low-density region in logarithmic scale on the $y$-axis: here the vertical dotted line at $\zeta=1$ serves as a guideline to highlight the onset of overlap, exceeds 1\% in packing fraction (horizontal line).
 (c) Example of three-microgel overlap occurring at $\zeta=2.02$ for surface-charged microgels, where these are almost faceted, with overlapping regions along the edges of contact surfaces (see enlargement on the right).
 (d) Example of many-microgel overlap occurring at $\zeta=3.76 $ for surface-charged microgels, where they almost fill the whole space and multiple overlaps appear at the vertices of contact surfaces. Here, protruding external chains are present (see enlargement on the right); still, at this $\zeta$ the volume fraction occupied by multiple overlaps is roughly only 1\% of the total overlap volume.
(e) True packing fraction $\phi-\phi_{ov}$ occupied by microgels (open symbols) and occupied fraction of the total volume $V_{tot}/L^3$ (filled symbols) as a function of $\zeta$; the black dashed line highlights $\phi=\zeta$.		}
		\label{fig:phi}
\end{figure*}

We now summarize the above results by reporting the evolution of the effective packing fraction, usually denoted as $\phi$, as a function of the nominal packing fraction $\zeta$. While the latter quantity is obtained from the individual particle size in the dilute limit (Eq.~\ref{eq:packfrac}), the former is  obtained by using $R_H(\zeta)$, i.e., the hydrodynamic size at each state point (Eq.~\ref{eq:packfrac2}), thus providing a better estimate of the volume effectively occupied by the particles. The parameter $\zeta$ is more widely used, since it is more easily accessible in experiments, while $\phi$ at large concentrations can be only determined by some clever estimation, as for example variational contrast scattering experiments~\cite{mohanty2017interpenetration}.
Fig.~\ref{fig:phi}(a) reports $\phi$ as a function of $\zeta$, confirming that $\phi$ coincides with $\zeta$ at low densities, but its slope decreases with increasing $\zeta$, in agreement with previous works~\cite{van2017fragility,gnan2019microscopic,nikolov2020behavior}.
The deviation between effective and nominal packing fraction occurs right at particle contact for neutral microgels, but slightly earlier for charged ones due to the anticipate shrinkage of their corona, as an effect of the long-ranged electrostatic interactions. However, for both systems, we find that $\phi$ rapidly exceeds 1.0, probably approaching a limiting value ($\gtrsim 1.5$) at very large $\zeta$. This suggests that, under very concentrated conditions, particles start to overlap, as also visualized in the snapshots, so that the packing fraction evaluated from individual microgels becomes an overestimation of the total packing fraction of the system.

To address this point, we next evaluate the overlap packing fraction $\phi_{ov}$. As explained in Methods, this is calculated as the fraction of volume occupied by more than a single microgel over the total one. 
With these calculations, we find that particle interpenetration starts at $\zeta \gtrsim 1.2$ for neutral and $\zeta \gtrsim 1.4$ for charged microgels.
This is highlighted in the inset of Fig.~\ref{fig:phi}(b), which displays the onset of the overlap in logarithmic scale. Clearly, for $\zeta \sim 1$ the amount of overlap is still well below 1\% of the total for neutral microgels, and even less for surface-charged ones.
These findings are again consistent with the fact that deswelling happens well before interpenetration takes place, in agreement with recent super-resolution microscopy experiments~\cite{conley2019}. Indeed, this study also reported that microgels first start to shrink at contact, while only later, roughly for $\zeta \gtrsim 1.1$, particle interpenetration becomes detectable.  In addition, our simulations also reveal that once overlaps appear, they grow quite significantly, so that $\phi_{ov}$ reaches values close to 0.3 for large $\zeta$.
The present results are also in agreement with neutron scattering experiments performed in ZAC, which detected the onset of interpenetration distinctly above contact~\cite{mohanty2017interpenetration}.
Importantly, in our simulations we are also able to distinguish between the types of overlaps that we observe. To this aim, we quantify the packing fraction of multiple overlaps, i.e. those occurring between more then two microgels in the same volume portion, also shown in Fig.~\ref{fig:phi}(b). They are illustrated in Fig.~\ref{fig:phi}(c) and (d), respectively, for three-body (typically along edges of facets) and many-body (typically at vertices of facets) overlaps. The contribution of such events to the total overlap volume fraction seems to be quite negligible. This suggests that interpenetration mostly occurs among two microgels in the explored range of packing fractions.

\begin{figure*}[ht]
	\centering
	\includegraphics[width=0.9\linewidth]{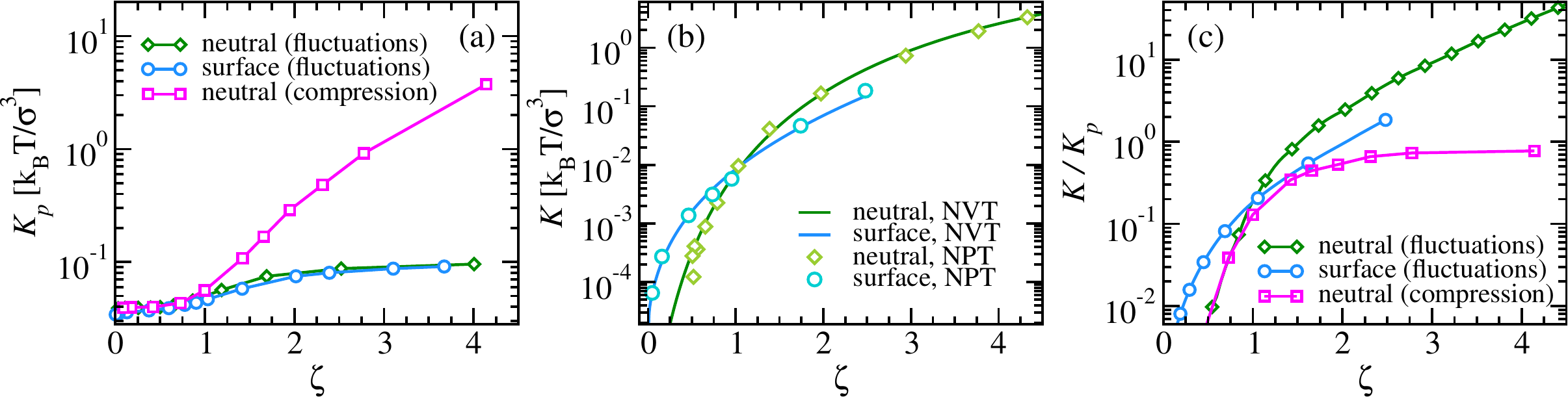}
	\caption{Bulk modulus as a function of $\zeta$: (a) for individual microgels $K_p$, calculated from fluctuations and from isotropic compression for neutral microgels; (b) for the whole suspension $K$ calculated from the derivative of the pressure with respect to volume and from NPT simulations; (c) ratio $K/K_p$. Lines are guides to the eye.}
	\label{fig:elasticity}
\end{figure*}

Finally, the effect of interpenetration can also be clearly seen by plotting the difference between the true individual microgels packing fraction and the overlap one, $\phi-\phi_{ov}$ in Fig.~\ref{fig:phi}(e).
The removal of the overlap contribution, that is of course double-counted in the calculation of $\phi$ assuming multiple overlaps give a negligible contribution, allows us to retain a clear plateau value for large $\zeta$.
This should ideally tend to 1, but we actually find an overestimation of about 10\%, due to the intrinsic difference between the geometric volume (used to calculate the overlap packing fraction) and the hydrodynamic one (used to calculate $\zeta$ and $\phi$). These two estimates are often used interchangeably, especially for soft particles, for which several definitions of volume can be thought, not generally identical to each other.
To remediate this problem,  we then simply calculate the total packing fraction of the system as the total volume occupied by all microgel beads $V_{tot}$ (based on the same mesh construction algorithm) divided by the volume of the box, also shown in Fig.~\ref{fig:phi}(e) and we correctly recover that it goes to $\sim 1$ for large packing fractions.
However, for small values of $\zeta$, this estimate no longer lies on the bisector, due to the difference with the hydrodynamic volume. Nonetheless, these calculations shed light on the way microgels occupy the space with increasing $\zeta$ and of the different mechanisms coming into play in the different regimes. Interestingly, the final saturation of the effective packing fraction roughly coincides with the end of the deformation in Fig.~\ref{fig:deformation}, so that in this final regime microgels have filled all the available space, reaching their maximum possible deformation. Hence,  at even higher $\zeta$, they just continue to further isotropically shrink (at fixed shape) and interpenetrate.

To summarize, both the present numerical analysis and the super-resolution experiments confirm that isotropic shrinking is the first effect taking place upon increasing $\zeta$, occurring right above particle contact. In a second step, deformation  and interpenetration, mostly in the form of two-body interactions, enter into play almost simultaneously. Finally, at very large $\zeta$, deformation stops and the physics of the suspension is completely dominated by further microgel shrinking and by the large amount of overlaps between the particles, also sometimes loosely referred as entanglements.

\subsection{Bulk modulus of the individual microgels and of the suspensions}
These observations are also mirrored in the behavior of  the elastic properties of the individual microgels and of the microgel suspension as a whole.
Similarly to previous works~\cite{lietor2011bulk,nikolov2020behavior}, we report the bulk modulus of the individual microgels $K_p$ as a function of $\zeta$ in Fig.~\ref{fig:elasticity}(a). This is evaluated in two ways: (i) from the volume fluctuations, similar to previous works~\cite{nikolov2018mesoscale,rovigatti2019connecting} and (ii) from isotropic compression of the microgels, as also done earlier both in bulk~\cite{rovigatti2019connecting} and at interfaces~\cite{vialetto2021effect,scotti2022softness}. The two methods should yield the same results, but this is true only in a limited range of packing fractions, namely $\zeta \lesssim 1.0$. Above this value, when microgels start to deform and interpenetrate, the results obtained with the two methods are remarkably different.  In particular, the estimates from volume fluctuations for both neutral and surface-charged microgels seem to saturate at finite, low value of $K_p\sim 0.1 k_BT/\sigma^3$. Instead, the data calculated via the compression for neutral microgels show a much more pronounced increase of $K_p$, which exceeds the other estimate by more than one order of magnitude. Comparing to previous data, experiments carried out in Ref.~\cite{lietor2011bulk} through osmotic compression of the microgels show a behavior of $K_p$ that is much more similar to data from compression. This suggests that the fluctuations route, amounting to spontaneous volume changes, does not probe the same fluctuations as those found when a physical compression, similar to the experimental one, is employed. Indeed, this method is based on the fluctuation-dissipation theorem and is strictly valid  for homogeneous closed systems only. Instead, in the present work, we applied the method to soft particles whose volume is not univocally defined, with different regions having different densities and reasonably different elastic properties~\cite{houston2022resolving}, and counterions and other microgels' beads in the same volume are not taken into account.

Next, we evaluate the suspensions bulk modulus $K$, that is shown in Fig.~\ref{fig:elasticity}(b). Again, we employ two different methods: directly from the system pressure $P$, as $K=-V(\partial P/\partial V)$ and from box volume fluctuations in NPT simulations.
We find, as expected, that the two methods yield identical results at all studied $\zeta$, showing a variation of $K$ of many orders of magnitude as a function of $\zeta$. Interestingly, we find that at low $\zeta$, surface-charged microgel suspensions are stiffer with respect to isotropic compression, but for $\zeta\gtrsim 1.0$ this trend is reversed and the neutral microgels become slightly more stiff, due to their larger degree of interpenetration at same $\zeta$.

Now looking at the ratio $K/K_p$ as a function of $\zeta$ in Fig.~\ref{fig:elasticity}(c), using $K_p$ obtained from compression, 
we find that it grows at first and then saturates for $\zeta \gtrsim 1.0$ at a constant value, that is found to be $\sim 1$. This happens exactly when deformation and interpenetration enter into play,  at which point the elastic modulus of the single particle increases of about one order of magnitude with respect to the dilute case, becoming comparable to the bulk modulus of the whole suspension. 
This suggests that at low $\zeta$, where microgels are not in contact, the corona of the microgels dominates the elastic response, while at high densities, microgels become all connected through interpenetration and they respond elastically as a whole.   For a matter of computational costs, we only show this for neutral microgels, but we have no reason to expect a different behavior for the surface-charged case. We also remark that Nikolov and coworkers~\cite{nikolov2020behavior}  as well as experiments in Ref.~\cite{lietor2011bulk} found a similar behavior, suggesting a universal trend for this observable and ruling out once again the ratio obtained from volume fluctuations, which continues to increase at large $\zeta$. 

\begin{figure*}[ht]
	\centering
	\includegraphics[width=0.9\linewidth]{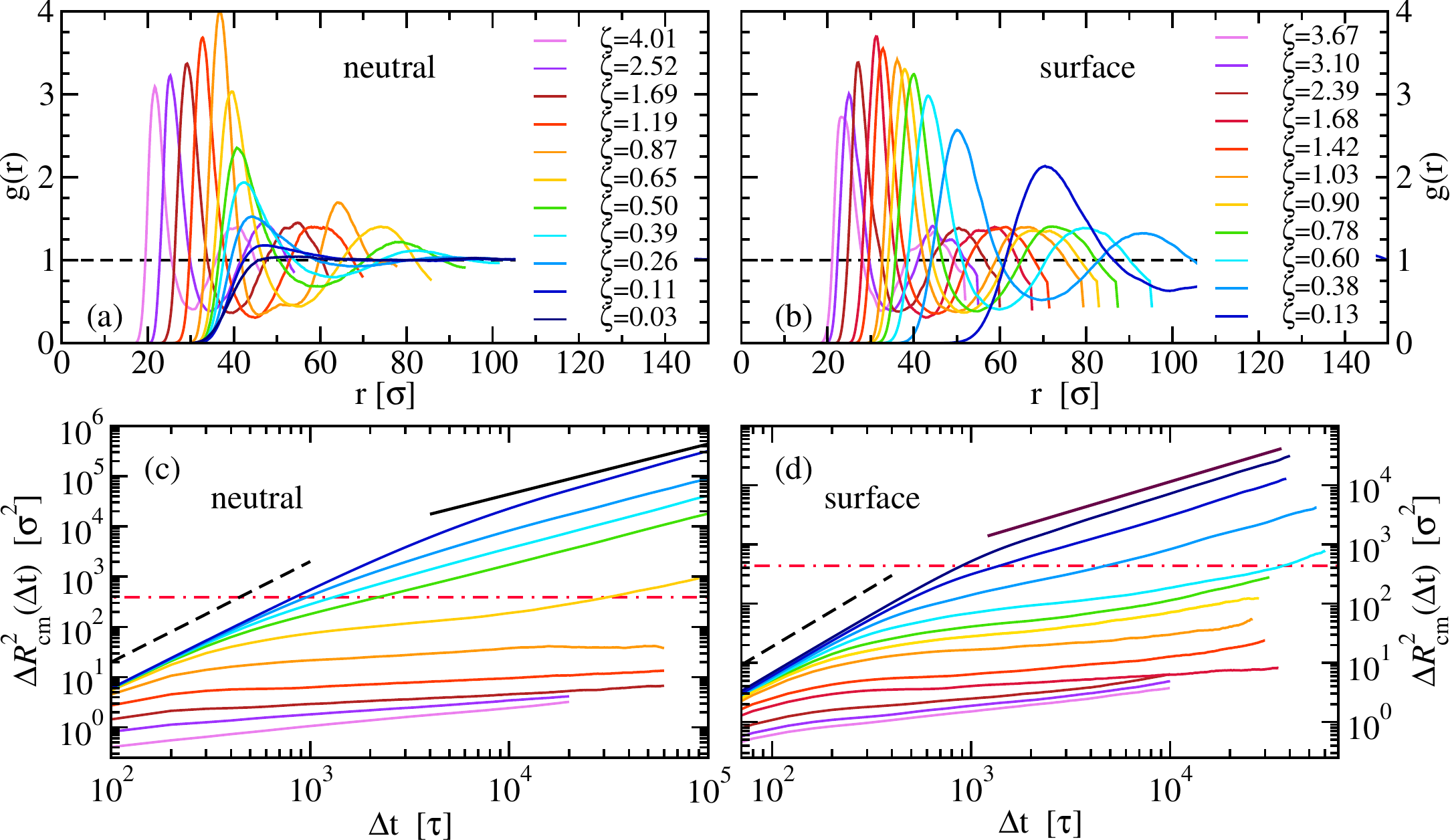}
	\caption{Radial distribution function as a function of $\zeta$ for (a) neutral microgels and (b) surface charged microgels; mean-squared displacement (MSD)  as a function of time for (c) neutral microgels and (d) surface charged microgels for different values of $\zeta$. The horizontal lines denote the squared hydrodynamic radius of the microgels at infinite dilution, while black dashed and continuous lines represent the ballistic and diffusive regimes, respectively. }
	\label{fig:gr-MSD}
\end{figure*}

\subsection{Collective behavior: radial distribution functions and mean-squared displacements}
Having discussed the single-particles properties in detail, as well as the influence of crowding by neighbouring particles on the internal structure of individual microgels, we now turn to examine the collective properties of the whole suspension. To this aim we analyze two key observables, representative of static and dynamic behavior, the radial distribution function $g(r)$ of the centers of mass of microgels and their mean-squared displacements (MSD), that are often discussed in experiments, particularly by microscopy ones.

We start by reporting $g(r)$ as a function of $\zeta$ in Fig~\ref{fig:gr-MSD} for both neutral (a) and surface-charged microgels (b). In both cases, the main peak shifts to smaller and smaller distances as $\zeta$ increases. However, strikingly, the height of the peak at first increases and then decreases at all  subsequent packing fractions for both types of microgels. This phenomenon, previously observed in confocal microscopy experiments~\cite{zhang2009thermal,paloli2013fluid} and also in their counterparts obtained by the study of the form factors~\cite{scotti2019deswelling}, is here referred as structural reentrance.

Next, we monitor the dynamics and report the MSD of the microgels centers of mass as function of time in Fig.~\ref{fig:gr-MSD} for neutral (c) and charged (d) microgels.
It can be observed that, for both cases, the  microgels are able to reach a fully diffusive regime at long times for low enough packing fractions. Then, their dynamics significantly slows down as $\zeta$ increases, approaching dynamical arrest.  This is indicated by the fact that microgels are no longer able to displace lengths larger than their average hydrodynamic radius (horizontal line in the figures) within the simulated time window. The onset of a glassy regime can be roughly estimated to take place for $\zeta \sim 0.8$
for neutral microgels, while it happens at even lower packing fractions for charged ones, due to the strong additional electrostatic interactions. 

Due to the large computational costs, we are not able to better characterize these high-density glassy states, nor to properly calculate the self-diffusion coefficient close to arrest in order to compare with available experimental data~\cite{philippe2018glass}. However, this behavior is qualitatively similar to that of the MSD measured with super-resolution microscopy~\cite{conley2017jamming}, wherein the plateau was found to dramatically decrease with increasing $\zeta$. However, we warn the reader that our system at these high packing fractions is out-of-equilibrium, and a true aging analysis should be performed, that is beyond current computational feasibility. In order to have a meaningful way to compare data at different packing fractions, we prepared the different state points following a similar protocol, without increasing the equilibration time with increasing $\zeta$, so that data in Fig.~\ref{fig:gr-MSD} are compared at the same waiting time. Despite the fact that we cannot be quantitative on the dynamics and on its $\zeta$-dependence, particularly to compare with the presence of a second regime at high $\zeta$ as proposed by Philippe and coworkers~\cite{philippe2018glass}, we can still put forward the robust observation that the MSDs are found to decrease monotonically with increasing $\zeta$, at any studied packing fractions, thus excluding a reentrant behavior of the dynamics in bulk microgel suspensions. 
Hence, for both studied microgel types, despite the presence of the structural reentrance, we do not detect any dynamical reentrance.

\subsection{Physical interpretation of the structural but not dynamical reentrance}

\begin{figure*}[ht]
	\centering
	\includegraphics[width=0.8\linewidth]{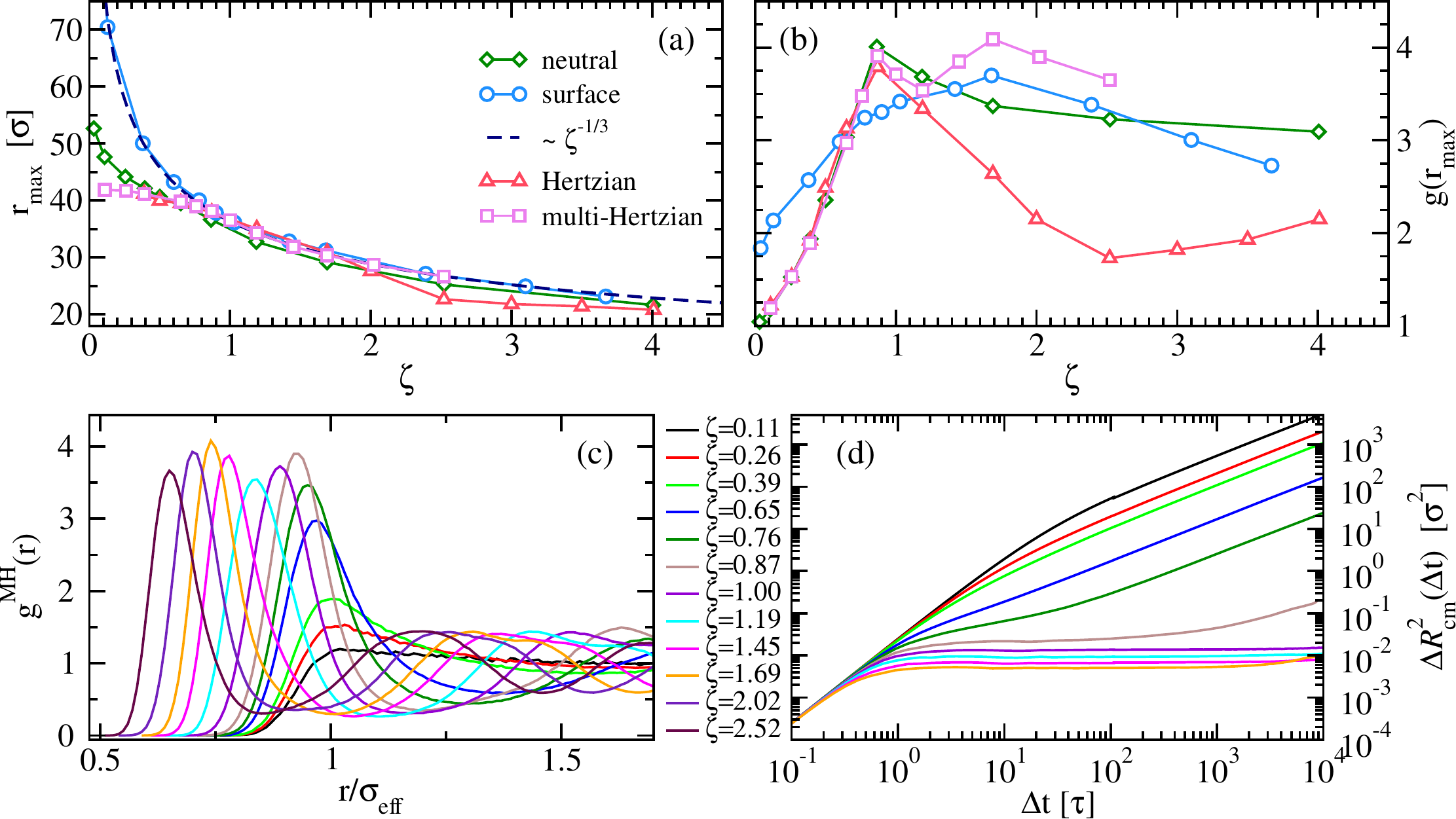}
	\caption{Evolution of the peak position $r_{max}$ (a) and of the peak height $g(r_{max})$ (b) as a function of $\zeta$ for simulations of neutral and charged microgels, as well as for a Hertzian and multi-Hertzian models (see text). MH simulations results for (c) radial distribution functions  $g^{MH}(r)$ and MSD (d) for different packing fractions.}
	\label{fig:gr-MH}
\end{figure*}

To shed light on the observed decoupling between statics and dynamics, we first discuss simple effective coarse-grained models. We start by considering the Hertzian model~\cite{zhang2009thermal,paloli2013fluid}, often invoked to describe microgels effective interactions in good solvent, because it is capable to describe the structural reentrance. 
We confirm this ability by comparing the monomer-resolved microgel data with those obtained by simulations of a simple Hertzian model, where $V_H (r) =\epsilon_H (1-r/\sigma_H)^{5/2}$. Similarly to previous analysis carried out in Refs.~\cite{mohanty2014effective,bergman2018new,rovigatti2019connecting}, the Hertzian diameter $\sigma_H$ is set equal to the hydrodynamic radius of the microgel for $\zeta=0$ and the packing fraction is determined accordingly. We then use the Hertzian strength $\epsilon_H$, which depends on the Young modulus of the microgels, as a fit parameter and find that $\epsilon_H \sim 470k_BT$ to match the data at low $\zeta$. Hence, we plot the position of the first peak $r_{max}$  and its height $g(r_{max})$ in Fig.~\ref{fig:gr-MH} (a) and (b), respectively. We find that for charged microgels, the behavior of  $r_{max}$   is consistent with an isotropic filling of the available volume $\sim \zeta^{-1/3}$,  at all packing fractions. For neutral microgels, this is roughly found only above contact, with some deviations at low $\zeta$. The Hertzian predictions are found to quite well reproduce the behavior of $r_{max}$ up to $\zeta\sim 2$. However, when we turn to look at $g(r_{max})$, we see that $V_H(r)$ can reproduce the simulations behavior only up to particle contact. For larger $\zeta$ values,  there is a much more dramatic decrease for the Hertzian spheres, than for the neutral microgels.  Interestingly, the Hertzian model further shows a minimum in $g_{max}$ at intermediate $\zeta$ ($\sim 2.5$), later followed by another increase of the peak height at ever larger packing fractions. This second increase of $g(r_{max})$ is not observed in monomer-resolved simulations, which instead display a much less pronounced decrease at large $\zeta$, with values always larger than those predicted by $V_H$. These results confirm what found in earlier works where the effective potential between two microgels was shown to deviate from the Hertzian model close to particle contact~\cite{rovigatti2019connecting}. In addition, simulations for $V_H(r)$ with the chosen repulsive strength are in the region where a dynamical reentrance is also present~\cite{berthier2010increasing}, at odds with the simulated microgels.

To grasp the physical meaning of the observed behavior, we recall why Hertzian spheres undergo structural and dynamical reentrance. Due to the soft elastic repulsion, it is more favourable for the particles  at high densities to become slightly overlapped, which in the monomer-resolved model is the effect given by particles shrinking, and deformation at larger packing fractions. In this way, they are able to free up some space in the available volume, so that at the same time, they can retain a more diffusive behavior. This implies that in the Hertzian model the structural reentrance is necessarily coupled to the  dynamical reentrance of the MSD. However, for realistic microgels, the mechanism at work is different,  since, even though particles are able to shrink, in doing so they have to retain the more external chains within the microgel, so that its inner part becomes stiffer and stiffer as $\zeta$ increases. To incorporate such an increase in the internal elasticity, it is useful to resort to yet another simple, two-body potential: the phenomenological multi-Hertzian (MH) model, previously employed to successfully describe the behavior of microgel mixtures~\cite{bergman2018new}. The MH effective potential amounts to the sum of three different Hertzian contributions to take into account corona-corona, corona-core (mixed) and core-core interactions, as discussed in Methods. 

The calculated radial distribution functions for the MH model, labelled as  $g^{MH}(r)$,  are reported in  Fig.~\ref{fig:gr-MH}(c) for various packing fractions. The peak position is now found to be in quantitative agreement with simulations in the whole investigated range, as also shown in Fig.~\ref{fig:gr-MH}(a),  while the peak height follows quite well the first initial decrease of the simulations up to $\zeta \gtrsim 1.5$, followed by a second increase. Hence, the MH model provides a noticeable improvement with respect to the Hertzian in terms of capturing the behavior of the structural reentrance. Most importantly, we then report the MSD of the multi-Hertzian model in Fig.~\ref{fig:gr-MH}(b),  for different packing fractions up to $\zeta=1.69$, because after this value we always find crystallization.  Strikingly, no dynamical reentrance is detected in qualitative agreement with the monomer-resolved simulations. Hence, the simple MH model is able to describe the main features of realistic microgels, despite the absence of other relaxing modes such as deformation and interpenetration, that generate many-body effective interactions and makes more difficult cage escaping events. Neglecting these complex effects, any simple coarse-grained model will necessarily fail in comparison to the true experimental data. Notwithstanding this, the MH offers an immediate physical interpretation of the decoupling between structural and dynamical reentrances due to the enhanced stiffness of the core, and even of the corona, as the microgels become more and more dense. In the case of charged microgels, it is more difficult to find a simple coarse-grained model, since the electrostatic repulsion needs to be described with further non-trivial degrees of freedom. Despite the fact that such a repulsion dominates over the elastic contribution at intermediate packing fractions, determining a delayed structural reentrance, the overall qualitative agreement with the behavior of neutral microgels confirms that the softness of the microgels coupled with their complex core-corona structure is the main responsible for the observed phenomenology.

\subsection{Onset of structural reentrance: a closer look}
We now try to grasp the physical origin of the structural reentrance. Starting with the simpler case of neutral microgels, we note that it takes place at $\zeta \sim 0.9$, as soon as particle deformation sets in and the Hertzian model starts to deviate from the prediction. Hence it is a combination of softness and deformation which gives rise to the phenomenology. The first aspect is also captured by the Hertzian model, that shares the reentrance feature with the monomer-resolve microgels only at a qualitative level. However, we can safely exclude any role of faceting and interpenetration in the phenomenon, since they only come into place at much larger $\zeta$.

\begin{figure}[ht]
	\centering
	\includegraphics[width=\linewidth]{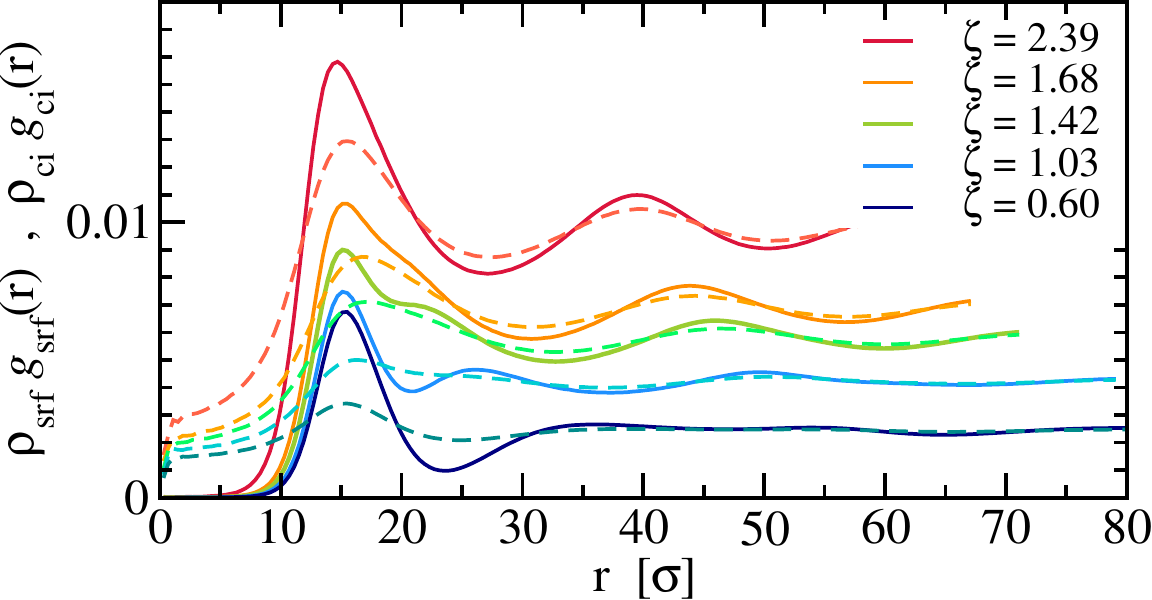}
	\caption{Radial distribution functions of charged beads $g_\text{srf}(r)$  (solid lines) and counterions $g_\text{ci}(r)$  (dashed lines), multiplied by their respective (identical) number densities $\rho_\text{srf}$ and  $\rho_\text{ci}$ , as a function of the distance from microgels center of mass, for selected values of $\zeta$ for surface-charged microgels.}
	\label{fig:gr-ions}
\end{figure}

Turning to surface-charged microgels, the situation is more intricate and the reentrance takes place at a larger packing fraction, around $\zeta\sim1.7$, when deformation is already in progress and the onset of isotropic shrinking just starts. To shed light on why this happens, we report the average radial density of charged beads $\rho_\text{srf}g_\text{srf}(r)$ and counterions $\rho_\text{ci}g_\text{ci}(r)$ as a function of the distance from the center of mass, in Fig.~\ref{fig:gr-ions}.
For small values of $\zeta$ we clearly distinguish a first peak in $g_\text{srf}(r)$, localized at the position of the microgel charged shell. This peak is intersected by the counterions distribution $g_\text{ci}(r)$, that is approximately located at the double layer of the microgels, and then followed by a second peak, which represents the average location of the external surface of the first neighbours. 
In this regime the mutual repulsion between the surfaces makes them stiffer, with respect to the neutral system at same $\zeta$.
A structural change of the charged distribution functions is then detected close to the packing fraction at which the reentrance of $g(r_\text{max})$ occurs: as visible from data in Fig.~\ref{fig:gr-ions}, for $1.4 \lesssim \zeta \lesssim 1.7$, the fusion among the first and second peak of $g_\text{srf}(r)$ clearly takes place. Thus, in this regime of packing fractions, the charged double layer disappears and the charged surfaces of neighbouring particles start to overlap, finally superimposing onto each other.
At even higher $\zeta$, we then mainly distinguish two kinds of regions in the system by the charged radial distributions:  denser and more positive cores, full of counterions, and the network of surface contact regions, that is instead  more negative due to microgel charges.
Importantly, a similar structural change was also detected in experimental studies, which highlighted the role of counterions screening and of their osmotic pressure on the swelling behaviour of pNIPAM microgels~\cite{gasser2019spontaneous}. Also, the properties of the counterion clouds of individual microgels have been shown to undergo dramatic changes when they start to overlap~\cite{scotti2016role}, in full agreement with the present work. 
The above evidence shows that the structural reentrance for charged microgels takes place in correspondence of the merging of the charged shells of neighboring particles. This suggests that for low packing fractions the interaction between the microgels is mainly dominated by the electrostatic repulsion, that keeps increasing while they approach each other until the charged surfaces merge and counterions are able to effectively screen them, while deformation and shrinking of the polymer network play a minor role and only start to dominate when screening become efficient. Hence, the onset of the structural reentrance also takes place in the presence of charges.

\subsection{Crystallization}
\begin{figure}[ht]
 	\centering
 	\includegraphics[width=0.9\linewidth]{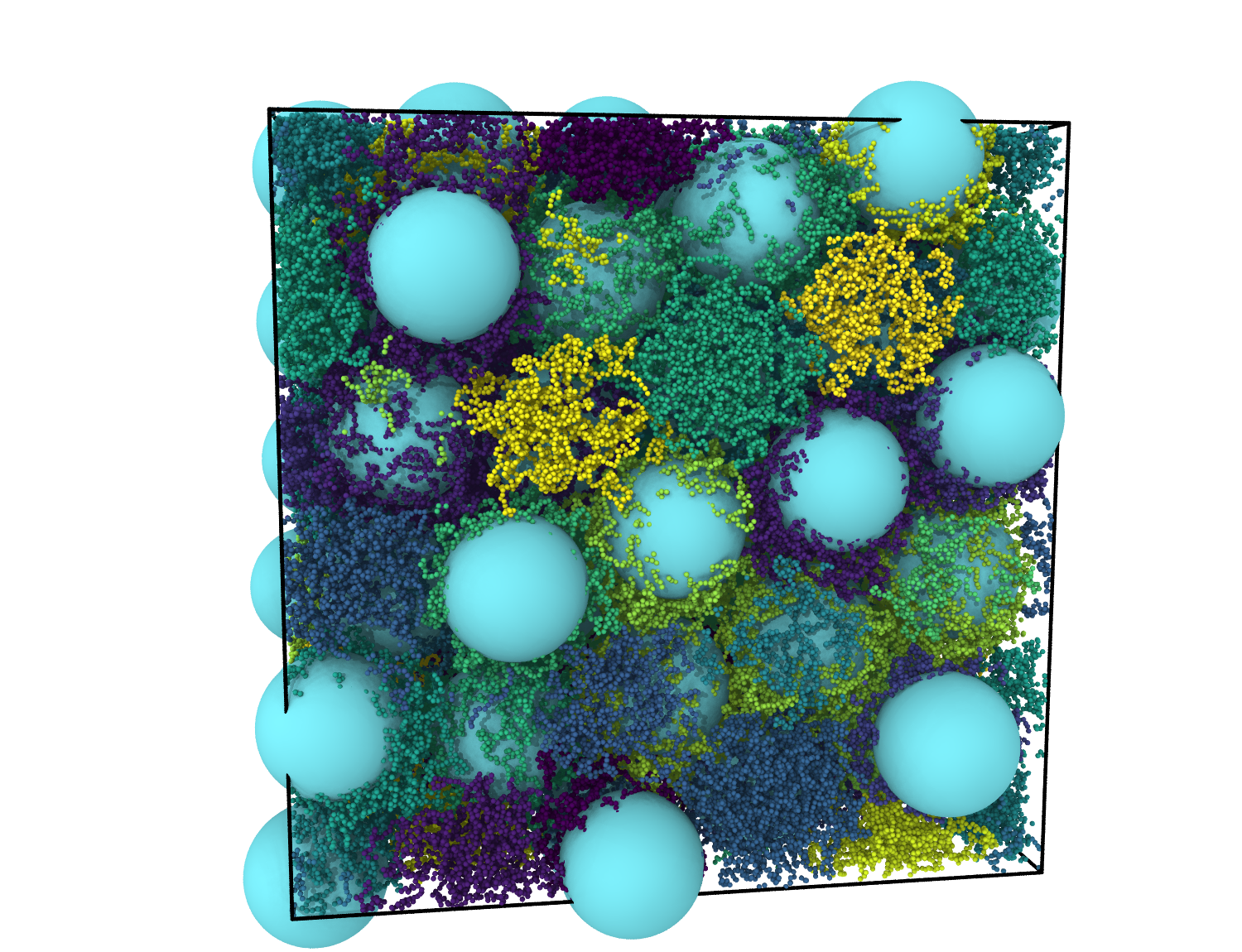}
 	\caption{Representative snapshot of the crystalline arrangement of  surface-charged $N=108$ microgels at $\zeta=2.39$. Monomer-resolved microgels are superimposed to a more coarse-grained representation in terms of effective (cyan) spheres of size $R_H(\zeta)$. This helps to visualize the cubic lattice formed by the centers of mass of such spheres.}
 	\label{fig:crystal}
 \end{figure}

Finally, we report that for both studied types of microgels, neutral and charged ones, we detect the onset of (at least partial) crystallization of the suspension at very high $\zeta$-values. Due to the underlying polydispersity of the employed microgels and to the small size of the sample, crystallization was not signaled by a sharp decrease of the energy (for which the bonded FENE contribution is always dominant), but is somehow visible from the splitting of the second peak of $g(r)$~\cite{paloli2013fluid} in some of the samples. For some systems where this splitting was observed, we ran longer simulations in order to be able to visualize the crystal.
One representative example is provided in the snapshot reported in Fig.~\ref{fig:crystal}, where it can be seen that the centers of mass of the microgels are found to occupy a cubic lattice.  Again, these findings indicate the realistic nature of our simulations, which display all the main features observed in experiments, including the onset of crystallization~\cite{paloli2013fluid}.

\begin{figure*}[ht]
	\centering
	\includegraphics[width=0.65\linewidth]{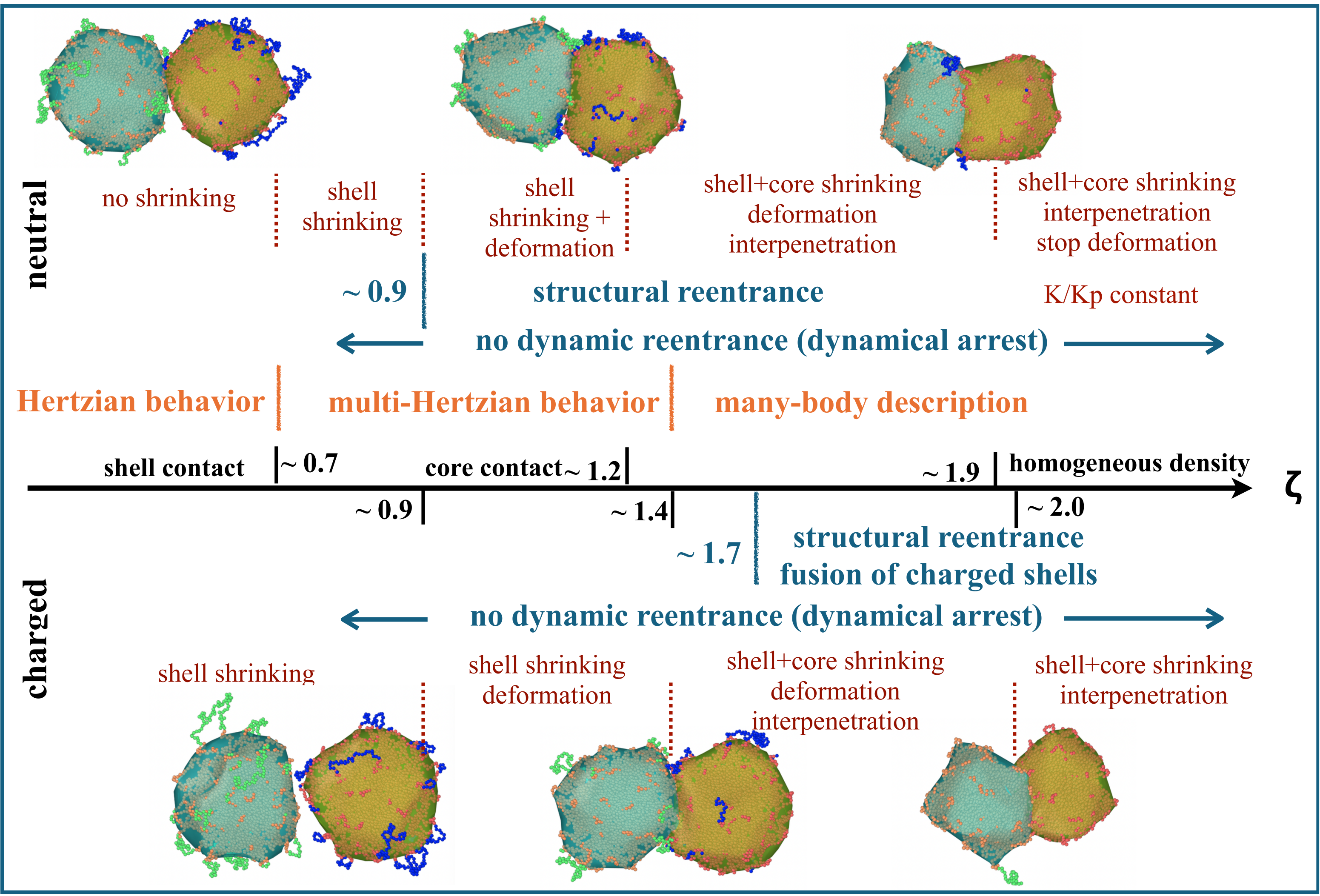}
	\caption{ Schematic representation of the different regimes of single-particle and collective behavior of neutral microgels observed in the present simulations as a function of nominal packing fraction $\zeta$. }
	\label{fig:scheffold}
\end{figure*}

\section{Discussion}
In this work we reported extensive simulations of suspensions of realistic microgels at varying packing fractions in the swollen regime. This model system is of high interest for the physics of glass transition,  because, at temperatures below the Volume Phase Transition, microgels are considered to be the prototype model of soft colloids. Therefore, their study at low temperatures allows to unveil important characteristics of the glass transition of soft particles in general. For a long time, in this regime, microgels have been thought to behave simply as Hertzian spheres, but several recent works have shown how such a pair potential is only adequate at low enough packing fractions, where particles do not strongly interact with each other. However, as soon as one reaches random close packing, which can be vastly exceeded in soft particles, an interesting region can be explored, where particles can change size, shape and interpenetrate, thus changing the rheological response of the suspension as a whole. Under these conditions,  more complex descriptions are needed. To this aim, novel microscopic insights have been provided recently by the advancement of super-resolution techniques, which have shed light on several aspects of concentrated microgel suspensions with single-particle resolution~\cite{conley2017jamming,conley2019,scheffold2020}. At a more global level, neutron studies with selective deuteration are also emerging as key experiments to unveil single microgel behavior under crowded environments~\cite{scotti2019deswelling}.

Complementary to these approaches, there is a strong need to perform state-of-the-art numerical simulations, which allow to follow the microgels at the microscopic level, both individually and collectively. In the present work, we were able to simulate up to 108 microgels, each made of about 5000 beads, which, following previous works, have been shown to possess an underlying structure similar to microgels synthesized in experiments~\cite{ninarello2019modeling}. Indeed, this is roughly the minimal size for which a meaningful core-corona topology, such as the one determined by experiments~\cite{stieger2004small}, can be achieved. The computational effort was even greater in the presence of charges, that we added to a fraction of beads similar to that of recent experiments~\cite{del2021two}, together with the corresponding counterions.
This aspect is often overlooked, due to the apparently negligible amount of charges present, that is due to the initiators of the polymerization reaction used in the synthesis, usually thought to be effectively screened by associated counterions.
However, more and more recent studies~\cite{scotti2016role,gasser2019spontaneous,del2021two,zhou2023measuring} pointed out that some features observed in experiments cannot be explained without accounting for electrostatic effects even for standard (e.g. non-copolymerized) pNIPAM microgels.
Hence, one of the main goals of the present study, that was never tackled before, was to perform a careful comparison between the behavior of neutral and charged microgel suspensions up to very high concentrations. 

For neutral microgels, the present simulations have enabled us to reveal with microscopic precision the different regimes taking place upon increasing packing fraction $\zeta$, which are found to be in remarkable agreement with super-resolution experiments by Scheffold and coworkers~\cite{conley2017jamming,conley2019}. 
In particular, the sequence in $\zeta$ is the same in both cases: shrinking, followed by deformation, then by interpenetration,  until finally deformation stops. These are graphically summarized in Fig.~\ref{fig:scheffold}. Noticeably, our estimates of the packing fraction values at which these take place  are in quantitative agreement with the experiments.  Instead, for surface-charged microgels we detect the same sequence of regimes, but shifted to higher packing fractions, due to the dominant role of electrostatic interactions at low and intermediate $\zeta$. This dominance comes to an end roughly where the charged shells of neighboring microgels merge together, a feature that we were able to detect by looking at the charged beads and counterions distribution functions, offering  for the first time a microscopic observation of a complex phenomenon, often invoked to explain experimental results~\cite{scotti2016role,gasser2019spontaneous}.
\\
In addition to  these results that are perhaps of specific interest only on the microgel community, our simulations also shed light on the often debated phenomena of structural and dynamical reentrances, that are features often associated to the behavior of soft spheres close to the glass transition. These concepts derived from early works on the Hertzian model, where they are coupled due to the softness of the underlying interactions. While it is now quite established that microgels behave as Hertzian particles only in a limited range of packing fractions, our monomer-resolved simulations of many microgels have allowed us to investigate these two reentrances in detail. We find that the structural reentrance occurs, but the dynamical one does not. This is in agreement with experimental observations, although a simultaneous study of both statics and dynamics was never reported within the same study for the same set of particles in experiments. Hence, a clear confirmation of this scenario was missing, which is now provided in the present work. Furthermore, we have identified the physical reasons for this behavior, namely that not only softness is at play. It is accompanied by deformation and also by the fact that as long as particles shrink, their internal elasticity varies due to the more external chains merging more and more, with the corona vanishing as $\zeta$ increases. This enhanced stiffness can be qualitatively captured by a simple modification of the Hertzian, namely the multi-Hertzian model, which shows the structural reentrance without the dynamical one, in agreement with the simulations. However, when deformation becomes dominant, even this modification fails and more complex models will have to be considered in the future to capture these many-body aspects.
\\
All this theoretical considerations strictly apply to neutral microgels, while for charged ones a starting point for a similar analysis could be a Hertzian complemented by a screened electrostatic repulsion~\cite{weyer2018concentration}, which is however beyond the scope of the present work.  However, the qualitative phenomenology is the same, confirming the generality of this behavior for core-corona microgels, independently on the amount of charged groups used in the synthesis. Remarkably, the onset of the structural reentrance in charged microgels takes place where the charged shells of neighbouring microgels fuse together, which offers a clear physical explanation of the phenomenon, due to the prevailing of soft over electrostatic interactions.  Importantly, the present work demonstrates that it is possible to observe  a structural reentrance, without necessarily having its dynamical counterpart, issues that are still debated in the literature, for microgels and other soft colloids.This implies that microgels can accommodate at increasing packing fractions, findings new ways of interacting which yield a different structural organization of their nearest neighbours. This is however not accompanied by the increase of available volume leading to a speed-up of the dynamics put forward by simple coarse-grained models, such as the Hertzian one. Instead, particles do reaccommodate, but in doing so they get more and more arrested, also due to the underlying interpenetration mechanism and related entanglements that increase the friction among them. Importantly, the mechanism of structural reentrance, albeit always present, is different in the presence of charges, because of the significant interactions between charged coronas which undergo a clear fusion in the interpenetration regime. It will be extremely interesting to try to detect this effect in experiments for example by means of selective labeling of the ionic groups.

In conclusion, the present results, being able to uncover all the many detailed specificities of microgel suspensions in swollen conditions, confirm the realistic nature of the employed microgel model and  offer the ideal framework to start tackling the rest of the phase diagram, where several questions still remain open and experimental observations are still scarce. In particular, there is a strong need to explore the temperature behavior where attractive interactions start to play a role, and also the different features that can arise by changing the topology of the microgels, e.g. working with ultra-low-crosslinked or hollow ones. These investigations will be the subject of future work.

\section{Methods}

We performed Molecular Dynamics (MD) simulations with $N=108$ microgels in a box with periodic boundary conditions. Each microgel is modelled as a connected network of $N_m\sim 5000$ beads, with a disordered topology, according to a recently proposed coarse-grained model~\cite{gnan2017silico,ninarello2019modeling}. To monitor size effects we also performed selected simulations with  $N=27$ microgels.  In all cases, beads interact with the Kremer-Grest potential, amounting to a steric repulsion plus a bond contribution between bonded neighbours. The first term is modeled by the Weeks-Chandler-Andersen (WCA) potential:
\begin{equation}
	\label{eq:wca}
	V_\text{WCA}(r)  =  
	\begin{cases}
		4\epsilon\left[\left(\frac{\sigma}{r}\right)^{12}-\left(\frac{\sigma}{r}\right)^6\right]+\epsilon & \quad \text{if} \quad r \le 2^{1/6}\sigma  \\
		0 & \quad \text{if}  \quad  r > 2^{1/6}\sigma
	\end{cases}
\end{equation}
where $\epsilon$ is the energy unit, $r$ identifies the distance between two monomers and $\sigma$ is the diameter of the monomer, that is the unit length of the simulation.
The bonding term is implemented through the Finitely Extensible Non-linear Elastic (FENE) potential:
\begin{equation}
V_{\text{FENE}}(r)  = 
-\epsilon k_F{R_0}^2\log\left[1-\left(\frac{r}{R_0\sigma}\right)^2\right], \quad r < R_0\sigma 
\end{equation}
where $R_0=1.5$ is the maximum bond distance and $k_F=15$ is a stiffness parameter determining the rigidity of the bond. 
We consider microgels with a crosslinker molar fraction $c\simeq 5.0\%$. Individual microgels are assembled using the procedure put forward in Ref.~\cite{ninarello2019modeling} and we use 27 independent topologies (replicated by 4 for the $N=108$ simulations) in order to reproduce the intrinsic polydispersity of microgel suspensions.

The presence of charged initiator molecules in the polymer network is taken into account in the model by having a fraction $f$ of network beads with a negative charge and interacting via the Coulomb potential~\cite{del2021two}:
\begin{equation}
	V_{\rm coul}(r_{ij}) = \frac{q_i q_j \sigma}{e^{*2} r_{ij} } \epsilon
\end{equation}
where $q_i$ and $q_j$ are the charges of the beads and $e^* = \sqrt{4\pi\varepsilon_0\varepsilon_r\sigma\epsilon}$ is the reduced unit  of charge, embedding the vacuum and relative dielectric constants, $\varepsilon_0$ and $\varepsilon_r$.
We consider $f=3.2\%$ with charges $q_i=-e^*$ and we analyse two different charge distributions as done in Ref.~\cite{del2021two}: (i) a random charge distribution, wherein charges are assigned to $fN_\text{m}$ monomers, randomly picked among non-crosslinker beads, and (ii) a surface charge distribution, wherein the same amount of charged beads are randomly spread only in the microgel external corona, i.e., randomly considering non-crosslinking beads whose average position from the center of mass is found at a distance larger than the gyration radius of the corresponding neutral microgel.
To preserve the overall electro-neutrality, we  insert an equivalent number of positive counterions with charge $q_i=+e^*$, which sterically interact among each other and with microgel beads through the WCA potential. Their diameter is set to $\sigma_c=0.1\sigma$ to avoid spurious effects from excluded volume~\cite{del2019numerical}.

The equations of motion are integrated through a stochastic algorithm for the canonical sampling~\cite{bussi2007canonical} (constant NVT ensemble) with an integration time-step $\Delta t = 0.002\tau$, where $\tau = \sqrt{m\sigma^2/\epsilon}$ is the reduced time unit and $m$ is the mass of the monomer bead.
All simulations are performed with  LAMMPS~\cite{LAMMPS,brown2011implementing,brown2012implementing} at fixed temperature $k_BT/\epsilon = 1.0$. Long-range Coulomb interactions are computed with the particle-particle-particle-mesh method ~\cite{deserno1998mesh}. Simulations are conducted for an equilibration time of $2\times 10^4\tau$ ($10^7$ timesteps), followed by a production run of $2\times 10^5\tau$ ($10^8$ timesteps) for systems with $N=27$, 
$5\times 10^4\tau$ ($2.5\times 10^7$ timesteps) for charged systems with $N=108$ and $10^5\tau$ ($5\times 10^7$ timesteps) for neutral systems with $N=108$.  We simulated the described systems for various values of the total simulation box volume $V$, to assess the effect of the concentration of microgels on collective and single-particle properties.

Additional simulations were performed for $N=5000$ particles interacting either with Hertzian ($V_H$) or multi-Hertzian model ($V_{MH}$) . The latter can be written as~\cite{bergman2018new},
\begin{equation}
V_{MH}(r)= V_H^{core}+V_H^{mid}+V_H^{corona},
 \label{eq:multi}
\end{equation}
where each term is a Hertzian potential for core-core, core-corona and corona-corona interactions. In particular, $V_H^{corona}$ coincides with the simple Hertzian describes in the text, so that with $\sigma_{\text{corona}}=\sigma_H$ and $U_{\text{corona}}=U_H$, while two additional parameters are introduced to describe each added Hertzian term, i.e. relative length scales $\sigma_{\text{core}}$, $\sigma_{\text{mid}}$, and their associated interaction strengths: $U_{\text{core}}$ and  $U_{\text{mid}}$. 
We fix the core diameter $\sigma_{\text{core}}=0.67\sigma_{\text{eff}}$, using its estimate obtained from the form factors fit in Fig.~\ref{fig:formfactors}, with $\sigma_{\text{mid}}=0.5(\sigma_{\text{core}} + \sigma_{\text{eff}}) = 0.835\sigma_{\text{eff}}$. 
We then vary the core-core Hertzian strength $U_{\text{core}}$ as a fit parameter in order to find the best match with the $g(r)$ calculated data for monomer-resolved simulations for neutral microgels. Here, the mixed interaction strength is again the average between the other two.  We find that the simulated data are in good agreement with the MH model when $U_{\text{core}}=3U_{\text{corona}}$.  

\subsection{Calculated observables}

We calculate the radial size of individual microgels through different quantities. An estimate of the overall extension of the core of the particles is given by the average radius of gyration:
\begin{equation}
\rg = \left\langle \frac{1}{N}\sum_{k=1}^{N}\sqrt{\sum_{i_k=1}^{N_{m,k}}\frac{\left(\vec{r}_{i_k} - \vec{r}_{cm,k}\right)^2}{N_{m,k}}} \right\rangle
\end{equation}
where $N_{m,k}$ are the number of beads of the $N$ microgels in the system, $\vec{r}_{i_k}$ is the position of the $i_k$-th monomer of the $k$-th microgel, whose center of mass (cm) has a position $\vec{r}_{{cm},k}$, and $\langle\cdot\rangle$ represents the ensemble average. This quantity represents the standard deviation of the radial density profile of microgels:
\begin{equation}
	\rho_M(r)= \left\langle \frac{1}{N}\sum_{k=1}^{N} \sum_{i_k=1}^{N_{m,k}} \delta (|\vec{r}_{i_k}-\vec{r}_{cm,k}|-r) \right\rangle.
	\end{equation}
	
Another estimate is given the hydrodynamic radius, which also accounts for their most exterior and low density region. 
To estimate $\rh$ in simulations, we adopt the method recently put forward in Ref.~\cite{del2021two}, where 
\begin{equation}
R_H = 2\left[ \int_{0}^{\infty}\frac{1}{\sqrt{(a^2+\theta)(b^2+\theta)(c^2+\theta)}}d\theta \right]^{-1}
\label{eq:Rh}
\end{equation}
and $a$, $b$, $c$ are the principal semiaxes of the gyration tensor of the surface mesh enclosing the microgel.
This is estimated with the Ovito python package~\cite{stukowski2009visualization} using the Alpha-Shape method~\cite{stukowski2014computational}, that is based on a Delaunay tassellation of the space occupied by microgels beads, then individuating filled and empty regions according to whether the radius of a circumscribing sphere is greater or smaller than a probing radius $r_p=8.0\sigma$, which in our case is a value that ensures the absence of inner holes at diluted conditions.  
The (nominal) packing fraction is thus defined as,
\begin{equation}\label{eq:packfrac}
	\zeta = \frac{4}{3}\pi R_{\text{H},0}^3 \frac{N}{V}
\end{equation}
where $R_{\text{H},0}$ is the hydrodynamic radius of the particles under most dilute conditions. This definition of the packing fraction is simply proportional to the number concentration and is usually used to report in experimental measurements, since it is easy to determine. However, due to the deformability of microgels, we also evaluate the effective volume fraction $\phi$, that is calculated by taking into account the actual hydrodynamic size $\rh$ of the particles at each state point:
\begin{equation}\label{eq:packfrac2}
	\phi = \frac{4}{3}\pi \rh^3 \frac{N}{V}.
\end{equation}

To probe the individual structure of microgels at different packing fractions, we calculate their form factors $P(k)$ with $k$ the wavenumber as
\begin{equation}\label{eq:Pk}
P(k) =\left\langle   \frac{1}{N}\sum_{j=1}^{N} \frac{1}{N_{m,j}}\sum_{n_j,l_j=1}^{N_{m,j}}  \exp \left[i\vec{k}\cdot (\vec{r}_{n_j}-\vec{r}_{l_j})\right] \right\rangle
\end{equation}
averaged over all microgels in the simulations. In order to compare with experiments, we then fit  $P(k)$ with a modified fuzzy sphere model model~\cite{ninarello2019modeling},
\begin{eqnarray}\label{eq:mod_fuzzy}
	P(k) &\propto& \left\{\left[ \frac{3\left(\sin(kR_c)-kR_c\cos(kR_c)\right)}{\left(kR_c\right)^3} \right.\right.\\
		 & +	 & \left.\left. s\left( \frac{\cos(kR_c)}{k^2R_c} - \frac{2\sin(kR_c)}{k^3R_c^2} - \frac{\cos(kR_c)-1}{k^4R_c^3} \right)\right] \right. \nonumber\\
		 &\times & \left. \exp\left(-\frac{\left(k\sigma_\text{surf}\right)^2}{2}\right) \right\}^2  +  \frac{1}{\left[1+\frac{D+1}{3}\xi^2k^2\right]^{D/2}} \nonumber
\end{eqnarray}
where the last modified Lorentzian term takes into account the network structure at large $k$~\cite{shibayama1992small}, based on the assumption that spatial correlations decay according to $r^{D-3}$,  where $D$ is the fractal dimension of the correlated domains and $\xi$ being the length over which concentration fluctuations are spatially correlated. In this way we obtain an estimate of the average core radius $R_c$ and of the average total radius $R_F=R_c+2\sigma_\text{surf}$, where $\sigma_\text{surf}$ denotes the extension of the corona.

To assess the structural changes occurring on single microgels upon increasing packing fraction, we additionally carried out calculations to evaluate variations of their shape.
To define a parameter that quantifies the deviation of microgels' shape from the spherical one, we construct the 
surface mesh through the aforementioned method, but we only consider the monomer beads that are found to be inside a certain radius, which we calculate as $l_h\equiv R_C+\sigma_\text{surf}$. Here $R_C$ and $\sigma_\text{surf}$ are obtained from the fit of the form factor in dilute conditions. In such a way, we can better define the surface of the particle to more accurately estimate its instantaneous shape by considering almost all its monomers except for the most external ones, which belong to the fluctuating outer chains and have a huge effect on the definition of the surface mesh. We justify our choice by considering that these dangling chains cannot be really detected in experimental systems through imaging techniques due to the very low contrast with the solvent~\cite{mohanty2017interpenetration,conley2019}. 
Following the method used in Ref.~\cite{conley2017jamming} to calculate the shape parameter, we thus use an analogue 3D quantity:
\begin{equation}
S_{p} = \frac{1}{N}\sum_{i=1}^{N}\frac{6\sqrt{\pi}V_i}{S_i^{3/2}}
\label{eq:deformation}
\end{equation}
where $S_i$ is the enclosing surface of each microgel. We note that according to this definition $S_{p} = 1$ for a sphere.
Our somehow arbitrary choice of $l_h$ are then validated \textit{a posteriori} by the results shown in Fig.~\ref{fig:deformation},  as discussed in the text.

Having defined a surface with an enclosed volume for each microgel, we then calculate the total overlap between different ones, which quantifies their overall degree of interpenetration. This is obtained as the fraction of total volume occupied by more than a single microgel:
\begin{eqnarray}
\phi_{ov} = \frac{V_\text{ov}}{V_\text{tot}}  \, &,&\quad V_\text{ov} = \left\langle\bigcup\limits_{i=1,j>i}^{N}V_i(l_h)\cap V_j(l_h)\right\rangle \, ,\nonumber\\
 & & V_\text{tot}=\bigcup\limits_{i=1}^{N}\left\langle V_i(l_h)\right\rangle
\end{eqnarray}
where $V_i(l_h)$ indicates the instantaneous volume of each surface mesh.
We use $\phi_{ov}$ as an estimate of the average amount of interpenetration of the system.

To estimate the bulk modulus of individual microgels, we perform two different calculations. First, we evaluate spontaneous volume fluctuations of the microgels in the concentrated suspensions, averaging the results over all  microgels, as
\begin{equation}
K_p=\frac{k_BT<V_i>}{<V_i^2> -<V_i>^2}
\label{eq:kp}
\end{equation}
where $V_i$ is calculated as before through the surface mesh of the microgels. Second, we run compression simulations by imposing a harmonic force on the microgel, $F(r)= -k (r-R)^2$ with $F(r)=0$ for $r<R$. Here, R is the radius of spherical confinement, $r$ is the distance of each monomer of the microgel from the center of the sphere and $k=10$. For each considered value of $R$, we run five independent simulations of different microgel topologies and average results over them. We then calculate $K_p$ again with Eq.~\ref{eq:kp} but within this constrained simulation and map the considered $R$ to the nominal packing fraction $\zeta$ by matching results to the fluctuations ones at low $\zeta$ where the two are identical. This method only works at not too small $R$ when we are in the linear regime. In addition, we have used it only for neutral microgels, because of the problem to treat the counterions in the charged case. The suspension bulk modulus is evaluated from the numerical derivative of the equation of state of our NVT simulations, i.e. $K=-V(\partial P/\partial V)$. In order to reduce statistical noise in the derivative, we also took the equilibrated configurations from the NVT simulations and simulated them at their equilibrium pressure. With these additional $NPT$ simulations, we could then directly estimate $K$ from the box fluctuations, analogously to Eq.~\ref{eq:kp}. In this way, we verified that the two approaches are identical and cleaned the statistical noise occurring at low $\zeta$.

To study the collective structure of the microgel suspension, we calculate the radial distribution function $g(r)$ of microgels centers of mass. In addition, we also calculate the radial pair distribution functions for charged monomers $g_\text{srf}(r)$ and for counterions $g_\text{ci}(r)$ both with respect to centers of mass, that are reported multiplied by their (identical) number densities $\rho_\text{srf}=\rho_\text{ci}=fN/V$, which by definition represent their average radial densities as a function of the distance from the center of mass of a given microgel. Finally, the dynamics of the system is monitored  through the calculation of the mean squared displacement of microgels' center of mass:
\begin{equation}\label{eq:msd}
	\Delta R_\text{cm}^2(\Delta t) = \overline{\frac{1}{N}\sum_{i=1}^{N}\left|\vec{R}_i(\Delta t)-\vec{R}_i(0)\right|^2}
\end{equation}
where $\overline{\cdot}$ indicates an average along the same trajectory. 

\section*{Acknowledgments}
We thank F. Camerin and L. Rovigatti for extensive discussions and methodologies developed throughout the years to simulate realistic microgels. We acknowledge the CINECA award under the ISCRA initiative, for the availability of high-performance computing resources and support. EZ acknowledges financial support from EU MSCA Doctoral Network QLUSTER, Grant Agreement 101072964 and from ICSC Centro Nazionale di Ricerca in High Performance Computing, Big Data and Quantum Computing, funded by European Union, NextGenerationEU - PNRR, Missione 4 Componente 2 Investimento 1.4.

\bibliography{bibliography}

\end{document}